\documentclass[lettersize,journal]{IEEEtran}
\usepackage{amsmath,amsfonts}
\usepackage{algorithmic}
\usepackage{algorithm}
\usepackage{array}
\usepackage[caption=false,font=normalsize,labelfont=sf,textfont=sf]{subfig}
\usepackage{textcomp}
\usepackage{stfloats}
\usepackage{url}
\usepackage{verbatim}
\usepackage{graphicx}
\usepackage{cite}
\usepackage{subfloat}
\usepackage{subcaption}
\usepackage{tabularx} % 引入 tabularx 宏包
\usepackage{ragged2e} % 引入 ragged2e 以支持换行时的左对齐
\usepackage{multirow}

\usepackage{amsmath,amssymb,amsfonts}
\usepackage{algorithmic}
\usepackage{textcomp}
\usepackage{hyperref}

\usepackage{booktabs}

\usepackage{subfloat}
\usepackage{enumitem}
\setlist[itemize]{leftmargin=*}
\usepackage{xcolor}
\definecolor{backred}{RGB}{255, 190, 190}
\definecolor{backblue}{RGB}{210, 230, 250}
\definecolor{verylightgray}{gray}{0.95}
\usepackage{tcolorbox}
\usepackage{color, colortbl}
\newcommand{\bestOS}[1]{#1}
\newcommand{\bestComm}[1]{#1}

\begin{document}

\title{FashionKG-RAG: Knowledge Graph-Enhanced Retrieval-Augmented Generation for Fashion Question Answering}

\author{Yujuan Ding, Linyin Luo, Shijie Wang, Xu Yuan, Yunshan Ma, Yi Bin, Wenqi Fan, Qing Li,~\IEEEmembership{Fellow,~IEEE}

\thanks{Yujuan Ding, Shijie Wang, Xu Yuan and Qing Li are with Department of Computing, Hong Kong Polytechnic University, Linyin Luo is with School of Computer Science and Engineering, Sun Yat-sen University; Wenqi Fan is with Deprtment of Computing and Department of Management and Marketing, Hong Kong Polytechnic University; Yunshan Ma is with School of Computing and Information Systems, Singapore Management University; Yi Bin is with School of Computer Science and Technology, Tongji University. }}% <-this % stops a space

\markboth{Journal of \LaTeX\ Class Files,~Vol.~14, No.~8, August~2021}%
{Shell \MakeLowercase{\textit{et al.}}: A Sample Article Using IEEEtran.cls for IEEE Journals}

\maketitle

\begin{abstract}
Fashion is a knowledge-intensive domain in which effective decision-making depends on integrating multiple types of knowledge. Although Large Language Models (LLMs) have transformed many areas, their application in fashion remains limited by hallucinations and weak domain specialization. Knowledge Graph (KG)-based Retrieval-Augmented Generation (RAG) offers a promising way to add structured knowledge to LLMs. However, existing fashion KGs are typically restricted to product-level attributes or item relations, and fail to capture the broader fashion ecosystem. To bridge these gaps, we propose \textbf{FashionEcoKG}, a comprehensive, domain-wide knowledge graph built with expert-level precision and professionalism. It is constructed through a three-stage agentic pipeline that extracts high-fidelity knowledge cores from authoritative textbooks and strengthens structural connectivity through cross-domain augmentation and generative expansion. To leverage this resource, we further develop \textbf{PG-RAG} (Pruning-Grounding RAG), a training-free framework designed to handle the conceptual density and linguistic noise of fashion queries. Specifically, we introduce a Dual-Granularity Path Re-Ranking (DGPR) module of two stages. The Pruning-based Semantic Ranking (PSR) module distills each query into a skeleton form to improve retrieval recall, while the Grounding-based Agentic Ranking (GAR) performs point-wise scrutiny of candidate paths against the original full query to ensure global relevance. Experiments on a curated fashion QA dataset show that PG-RAG effectively leverages FashionEcoKG to improve retrieval and answer accuracy, outperforming both non-RAG and existing KG-RAG baselines.

\end{abstract}


\section{Introduction}
Fashion is a cornerstone of the global economy and a critical application domain for Artificial Intelligence (AI)~\cite{oh2025fashion,liu2023toward}. While often perceived as a field driven by subjective aesthetics, fashion is fundamentally knowledge-intensive: successful decision-making relies on accumulated expertise spanning design, production, marketing, and communication~\footnote{https://www.mckinsey.com/industries/retail/our-insights/state-of-fashion\#/}. Current computational fashion research primarily focuses on visual perception, generation, and outfit-related tasks such as item retrieval~\cite{wang2023fashionklip} and recommendation~\cite{zhan20213}. However, the more fundamental and knowledge-intensive Question Answering (QA) task remains significantly underexplored~\cite{wang2023fashionvqa, ding2024fashionregen}.

\begin{figure}[t]
    \centering
    \includegraphics[width=1\linewidth]{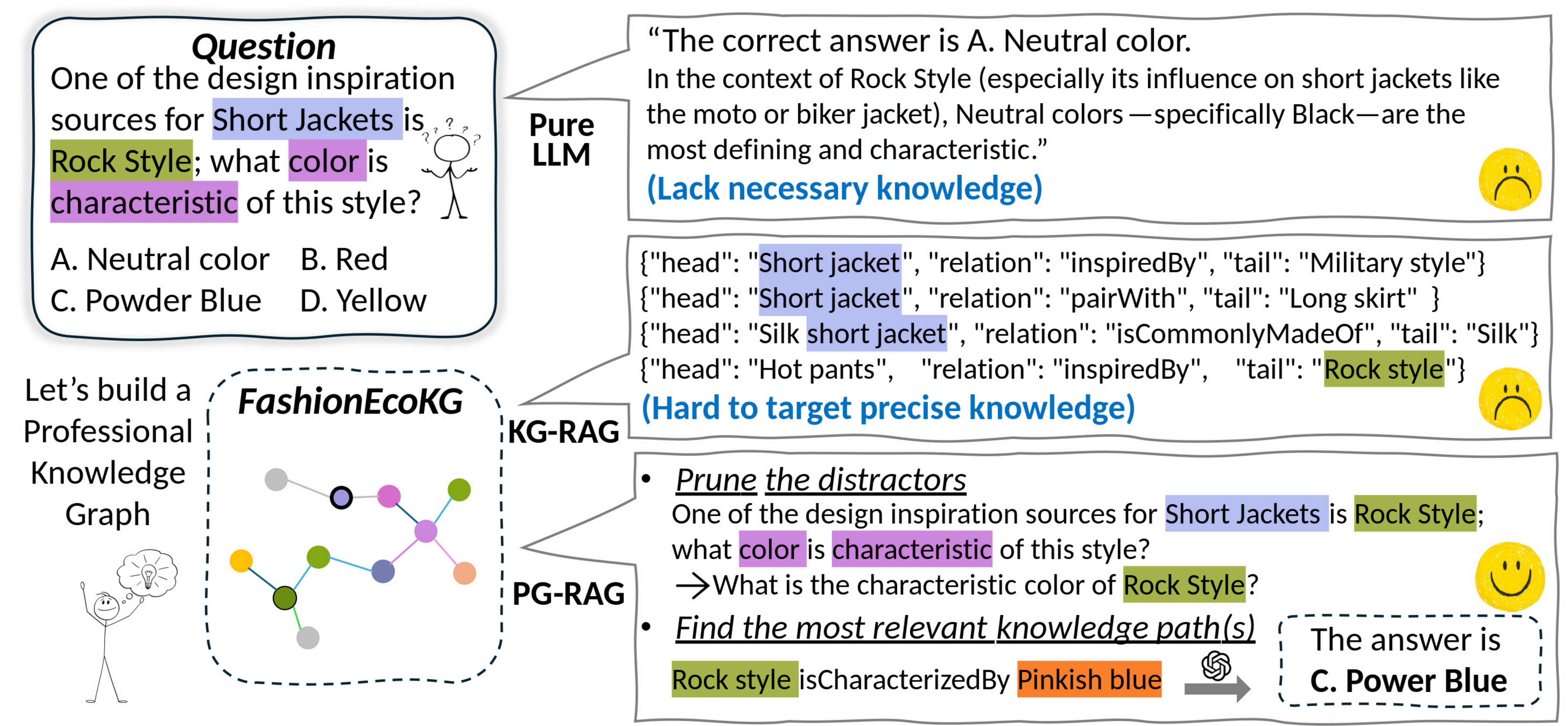}
    \caption{Illustration of knowledge-intensive fashion QA task addressed by various solutions. Off-the-shelf LLMs lack the necessary fashion knowledge for accurate responses. Even with well-structured knowledge graphs, naive KG-RAG methods struggle to precisely target relevant information due to the interference from query noise. In contrast, our 
    PG-RAG facilitates effective knowledge retrieval by pruning distractors and extracting the most relevant knowledge path(s).}
    \label{fig:task}
\end{figure}
Even though LLMs have demonstrated extraordinary capacity~\cite{OpenAIGPT5,DeepSeekAI2025DeepSeekR1IR} in general-domain tasks, they still face hallucination problems and struggle with specialized queries due to insufficient domain knowledge. As the example illustrated in ~\autoref{fig:task}, answering fashion questions with high conceptual density is challenging for general-purpose LLMs. Although Retrieval-Augmented Generation (RAG)~\cite{fan2024survey,gao2023retrieval} incorporates external references, standard flat text-based methods fail to leverage underlying structural knowledge. By neglecting relations between concepts separated across multiple pages, these retrievers provide fragmented information lacking long-range dependencies, which is insufficient for complex or deep reasoning~\cite{xu2024learning}. 

To address these limitations, Knowledge Graph (KG) integration offers a promising solution by providing structured domain intelligence without costly fine-tuning~\cite{liao2026enhancing}, with successful applications in other knowledge-intensive domains~\cite{matsumoto2024kragen,qi2026dual,gao2025large,shang2025personalized}. However, general-purpose KGs~\cite{auer2007dbpedia} often lack the specialized details required for advanced fashion reasoning. While fashion-specific KGs have been developed to fill this gap~\cite{jia2020fashionpedia}, they remain constrained by narrow focuses: product-level attributes for e-commerce~\cite{barroca2022enriching,jia2023kg,ding2023modeling}, basic relational patterns between items~\cite{zhan20213,wang2023fashionklip}, or transient social media trends~\cite{ma2019and,yuan2023multimodal}. Although these product-focused KGs may be effective and useful to enhance tasks such as recommendation, they are too narrow to capture the broader \textbf{Fashion Ecosystem}, which encompasses critical dimensions beyond fashion products, including business strategy, cultural history, and production science. Furthermore, because these resources are typically built on large-scale observational data rather than authoritative expert knowledge, they lack the professional rigor necessary for deep, knowledge-intensive analysis in the fashion domain. 

To close these gaps, we propose \textbf{\textit{FashionEcoKG}}, a comprehensive domain-wide knowledge graph. Unlike previous works that rely on unstructured web data, we utilize LLM agents to anchor our construction process in authoritative fashion textbooks, ensuring the fidelity and professional quality of the resulting knowledge. While LLMs show potential for automated KG construction~\cite{han2024pive,yang2025graphusion,chen2024sac,gao2025large,feng2025llmkg+}, building a high-quality expert-level graph from such dense academic sources presents unique challenges. 
% guarantees high precision, it 
Specifically, restricting our source to canonical references initially yields an inherently narrow core with limited situational coverage. Furthermore, because textbooks often present concepts through high-level exemplification without explicitly mapping every relational dependency, the initial extraction often results in fragmented or isolated entities. 
To achieve both domain expertise and structural connectivity, we propose a novel three-stage hierarchical pipeline based on LLM agents that evolves a sparse textbook core into a larger and more comprehensive network.  
First, LLM agents extract concise entities and triples from canonical references to establish a high-quality foundation. Next, we augment this core with general-domain KGs to bridge isolated entities and ensure structural connectivity.
Finally, we expand the graph's coverage and further enhance its relational structure using validated, LLM-generated knowledge. 

As illustrated in ~\autoref{fig:task}, this work focuses on knowledge-intensive fashion QA, where queries typically exhibit high conceptual density and frequently span diverse sub-domains. This multifaceted nature inevitably introduces semantic noise, as mentions or concepts that appear relevant may actually serve as distractors, complicating both the retrieval and generation phases. This distraction may be more amplified in a KG-RAG pipeline than in a pure LLM due to error accumulation. 
To address these challenges, we propose \textbf{\textit{PG-RAG}} (Pruning-Grounding RAG), a KG-enhanced framework utilizing a multi-stage path retrieval and re-ranking process. Unlike conventional entity-based reasoning, our approach organizes the FashionEcoKG into multi-hop paths to capture the structural topology. 
After an initial entity-based coarse retrieval, we introduce Dual-Granularity Path Re-Ranking (\textbf{\textit{DGPR}}), an architecture that re-ranks the candidate knowledge paths to reconcile retrieval breadth with precision.
This dual-stage design begins with Pruning-based Semantic Ranking (PSR), which distills the conceptually dense fashion query into a simplified skeleton to neutralize combinational distractors. 
PSR serves as a necessary filter to isolate core intent and ensure a high-recall semantic search without interference from linguistic noise. 
This significant reduction in candidates subsequently enables Grounding-based Agentic Ranking (GAR), where an LLM agent performs point-wise scrutiny against the original full query. 
% By verifying that retrieved reasoning chains logically satisfy granular, expert-level constraints, GAR provides the final structural grounding necessary for precise information retrieval in expert-intensive vertical domains. 
% GAR acts as a final check, ensuring that the chosen knowledge paths logically match every detail of the question to provide a precise answer.
Through this \textit{Pruning-to-Grounding} transition, PG-RAG ensures that the final output is anchored in semantic intent and also considers global context and details. 
% We conduct extensive experiments on a curated knowledge-intensive FashionQA dataset. Experimental results demonstrate the effectiveness of our proposed PG-RAG through the comparison between no-RAG and other existing KG- or Graph-RAG methods.  It is also proved that our FashionEcoKG is critical to help with the QA accuracy. 

Our work addresses the limitations of existing fashion QA models through the following technical contributions:
\begin{itemize}
    % \item We contribute FashionEcoKG, a comprehensive, expert-anchored Knowledge Graph sourced from authoritative references. It uses a three-stage pipeline (Distill, Align, Expand) to evolve sparse academic data into a dense, professional network. Unlike previous graphs that focus solely on product attributes or social media metadata, FashionEcoKG provides an expansive coverage of the fashion ecosystem. 
    \item We introduce \textbf{\textit{FashionEcoKG}}, a comprehensive, domain-wide Knowledge Graph anchored by authoritative textbooks. By employing an effective three-stage construction pipeline (Extraction, Augmentation, Expansion), we evolve sparse academic concepts into a professional network that captures the broader fashion ecosystem knowledge beyond simple product-centric relations.
    % or social media observations.  
    \item We propose \textbf{\textit{PG-RAG}} (Pruning-Grounding RAG), a KG-enhanced framework featuring Dual Granularity Path Re-Ranking (\textbf{\textit{DGPR}}). This multi-stage re-ranking architecture reconciles retrieval breadth with reasoning precision by transitioning from Pruning-based Semantic Ranking (PSR) for noise neutralization to Grounding-based Agentic Ranking (GAR) for logical and detailed verification against the original query.
    \item We evaluate our method through extensive experiments on a fashion QA dataset with a curated set of multi-type, multi-complexity domain-specific questions. The results show that PG-RAG significantly outperforms traditional non-RAG, text-based RAG, and existing KG-RAG baselines, proving its superior ability to leverage structured domain intelligence for better QA precision. 
\end{itemize}

\section{Related Work}
This section reviews related studies on Knowledge Graph-based RAG, Fashion KG, and RAG for Fashion.

\subsection{Knowledge Graph-based RAG}
RAG is widely adopted in LLMs to mitigate issues such as outdated knowledge and hallucinations~\cite{fan2024survey,ni2025towards}. By dynamically integrating external information with parametric knowledge, RAG has been proven effective for knowledge-intensive tasks across diverse domains~\cite{matsumoto2024kragen,dutta2024rar,gao2025large,yuan2025mkg}.
However, conventional RAG methods primarily rely on unstructured knowledge representations and often overlook logical relationships among knowledge elements, limiting their capacity for complex or multi-hop reasoning~\cite{zhang2025survey}.

To address this limitation, Knowledge Graph-based RAG (KG-RAG) has recently emerged as a paradigm for augmenting LLMs with structured knowledge. 
Based on how graph structures are utilized, KG-RAG approaches can be broadly categorized into three types~\cite{zhang2025survey}.
(1) \emph{Knowledge-based KG-RAG} treats graphs as structured knowledge carriers to enable efficient retrieval and multi-step reasoning. Early works focus on reasoning over existing KGs, such as KAPING~\cite{baek2023knowledge} and ToG~\cite{sun2023think}. Another line of research converts unstructured corpora into structured graphs. For example, GraphRAG~\cite{edge2024local} and LightRAG~\cite{guo2025lightrag} leverage LLMs to automatically construct attributed graphs with enriched node and edge semantics, while StructRAG~\cite{li2025structrag} explores richer structures by introducing five task-specific graph types.
(2) \emph{Index-based KG-RAG} uses graphs as indexing structures to organize and retrieve raw text chunks for LLM inference. GNN-ret~\cite{li2025graph} constructs passage-level graphs based on structural and keyword similarities to support general-purpose retrieval. KGP~\cite{wang2024knowledge} extends this approach by building question-aware indexing graphs that capture lexical, semantic, and entity-level relationships.
(3) \emph{Hybrid KG-RAG} utilizes graph structures both as carriers of knowledge and as indexing tools~\cite{zhu2025knowledge,zhuang2025linearrag}. A representative approach is LinearRAG~\cite{zhuang2025linearrag}, which builds a lightweight hierarchical graph over entities, sentences, and passages, enabling passage retrieval through local semantic bridging and global importance aggregation.

\begin{figure}[t]
    \centering
    \includegraphics[width=\linewidth]{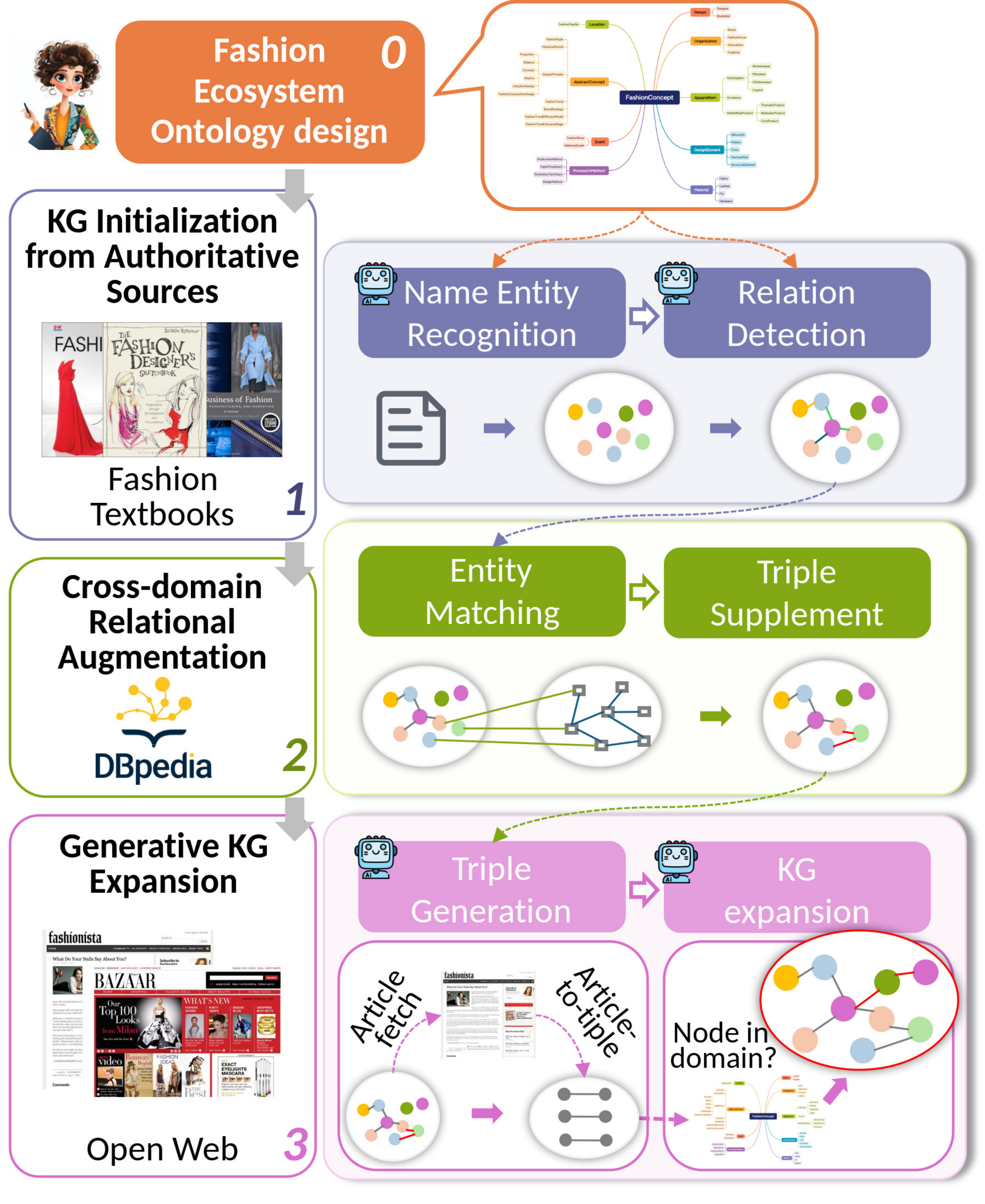}
    \caption{FashionEcoKG construction pipeline. The initialization stage extracts a concise and professional knowledge core from authoritative textbooks, which then evolves into a network with better connectivity and coverage through two phases of augmentation.}
    \label{fig:fashionEcoKG}
\end{figure}

\begin{figure}[t]
    \centering
    \includegraphics[width=1\linewidth]{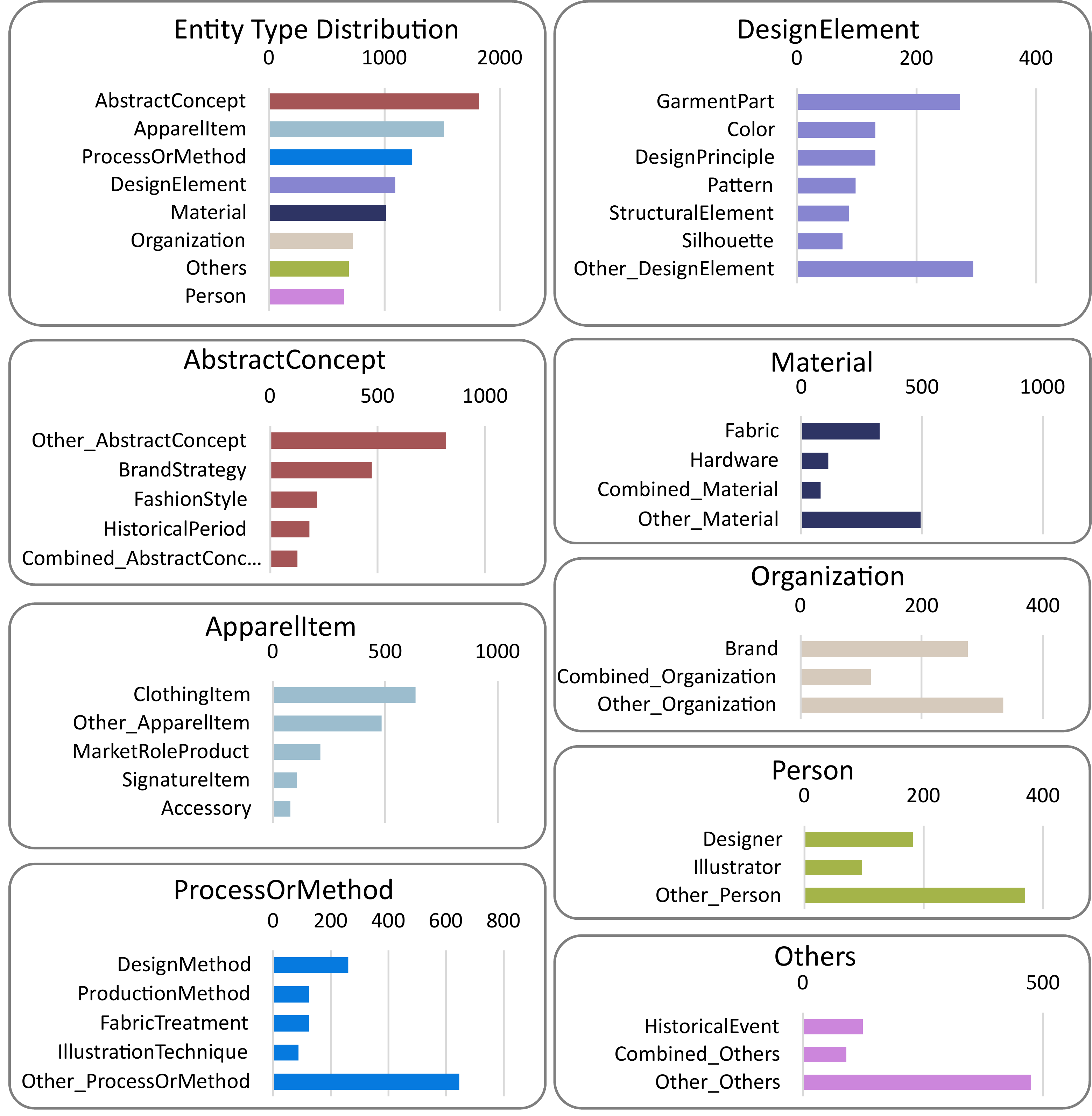}
    \caption{Entity type distribution of FashionEcoKG. The top-left panel illustrates the primary entity types, while the remaining subplots provide a granular breakdown of secondary entity types for each respective primary type.}
    \label{fig:ontology}
\end{figure}

\subsection{Fashion KG and RAG for Fashion}
Knowledge graphs in the fashion domain have traditionally focused on product-centric tasks. 
% have been applied in the fashion domain primarily to support product-centric tasks. 
Early work, such as FashionPedia~\cite{jia2020fashionpedia}, focuses on fine-grained visual understanding of apparel, enabling product classification and attribute detection.
% through structured representations of apparel attributes and relations. 
Similarly, \cite{zhan20213} constructs a fashion KG based on item- and outfit-level attributes for outfit recommendation. 
Recent studies incorporate multimodal information into fashion KGs. FashionKlip\cite{wang2023fashionklip} propose a multimodal fashion KG in which entities represent physical product attributes and edges encode semantic similarity, with a focus on e-commerce applications. Another line of work builds fashion KGs from social media data~\cite{ma2019and,yuan2023multimodal} to capture user behavior patterns and fashion trends. 
% Despite differing data sources, 
While effective for recommendation and retrieval~\cite{barroca2022enriching,yan2019differentiated,jia2023kg,ding2023modeling}, these methods are restricted to product- or outfit-level concepts. 
% As a result, existing fashion KGs are effective for downstream tasks such as recommendation and retrieval
% in e-commerce settings
% ~\cite{barroca2022enriching,yan2019differentiated,jia2023kg,ding2023modeling}. 
They largely neglect broader fashion knowledge, including cultural, historical, business, and design-related aspects, limiting their utility for complex domain reasoning.

Motivated by advances in KG-RAG, recent work explores integrating fashion KGs with LLMs. For instance, Fashion-RAG~\cite{sanguigni2025fashion} and FITMag~\cite{han12025fitmag} combine multimodal LLMs with graph-structured knowledge and social signals for fashion content generation. 
VestiMe~\cite{madurovestime} and VisioRAG~\cite{balachandran2025visiorag} leverage multimodal retrieval to support conversational exploration and recommendation. Despite these developments, the fundamental knowledge-intensive question answering task remains significantly underexplored in the fashion domain. 
Existing models~\cite{wang2023fashionvqa,xu2025itemrag} can identify a garment or garment-relevant concepts but often fail to abstract concepts, such as a designer’s underlying inspiration or the strategic rationale of a style. 
Such limitations hinder fashion AI from achieving expert-level decision-making and from addressing complex queries that require long-range reasoning across the broader fashion ecosystem.

\section{FashionEcoKG Construction}
\begin{table}[t]
\caption{Prompts for FashionEcoKG Initialization.}
\label{tab:KG_initialization_prompt}
\centering
\footnotesize
\begin{tcolorbox}
[
  title=NER Prompt,
  boxrule=0.6pt,
  arc=2pt,
  left=2pt,right=2pt,top=2pt,bottom=2pt,
  fonttitle=\bfseries
]

\textbf{System prompt:}
I will provide a section about the ontology design of the fashion knowledge graph and a chapter of content from a fashion-related book. Your task is to perform the NER task based on this content.

Requirements:
\begin{enumerate}
\setlength{\itemsep}{0pt}\setlength{\parsep}{0pt}\setlength{\topsep}{0pt}
    \item List all results as completely as possible, in Chinese.
    \item Do not include categories that are not in the ontology content.
    \item Entities should be specific instances rather than general categories and must not duplicate the existing entries in the ontology.
    \item Output as a JSON list in the format {{``name'': ``entity name'', ``label'': ``entity category''}}, where label includes all hierarchical categories separated by ``;''. Output only the result directly without explanation.
\end{enumerate}

\textbf{User prompt:}
\begin{enumerate}
\setlength{\itemsep}{0pt}\setlength{\parsep}{0pt}\setlength{\topsep}{0pt}
    \item Ontology Design: data[``ontology'']
    \item Text Content: data[``text'']
    \item Begin Output:
\end{enumerate}

\end{tcolorbox}

\begin{tcolorbox}
[
  title=RD Prompt,
  boxrule=0.6pt,
  arc=2pt,
  left=2pt,right=2pt,top=2pt,bottom=2pt,
  fonttitle=\bfseries
]
\textbf{System prompt:}
I will provide content related to the fashion domain knowledge graph, including a chapter from a fashion-related book, the corresponding NER results, and the designed relations. Your task is to perform the relation extraction task based on this content. You should identify as many entity relations as possible, where entities are from the provided NER results and relations are from the provided relation design. The output format should be triples.

\textbf{User prompt:}
\begin{enumerate}
\setlength{\itemsep}{0pt}\setlength{\parsep}{0pt}\setlength{\topsep}{0pt}
    \item NER Results: data[``NER'']
    \item Relation Design: data[``ontology'']
    \item Book Content: data[``book content'']
    \item Begin Output:
\end{enumerate}

\end{tcolorbox}
\end{table}

Our construction methodology follows a divergent yet seeded approach, designed to overcome the inherent trade-offs between knowledge authority and graph connectivity. We propose a Knowledge Initialization and Augmentation pipeline that transitions from a condensed, textbook-based core to a deep and large-scale knowledge network, as shown in ~\autoref{fig:fashionEcoKG}.

\begin{table}[t]
\caption{Prompts for FashionEcoKG Expansion.}
\label{tab:KG_expansion_prompt}
\centering
\scriptsize
\begin{tcolorbox}[
  title=Fetch articles prompt,
  boxrule=0.6pt,
  arc=2pt,
  left=2pt,right=2pt,top=2pt,bottom=2pt,
  fonttitle=\bfseries
]

\textbf{System prompt:}
I will provide a specific fashion entity keyword and its associated category label (context). Your task is to convert this keyword into a highly effective search query to retrieve professional content.

Requirements:
\begin{enumerate}
\setlength{\itemsep}{0pt}\setlength{\parsep}{0pt}\setlength{\topsep}{0pt}
    \item The goal is to find high-quality articles, fashion magazines, or academic papers.
    \item  Use the ``Label'' only as context to understand the domain (e.g., distinguishing a brand from a material), but do NOT clutter the search query with generic domain words unless necessary for disambiguation.
    \item  The query must focus primarily on the ``Keyword'' itself.
    \item  Output ONLY the raw query text string. Do not include quotes, explanations, or any other text.
    \item  The query should be in English.
\end{enumerate}

\textbf{User prompt:}
\begin{enumerate}
\setlength{\itemsep}{0pt}\setlength{\parsep}{0pt}\setlength{\topsep}{0pt}
    \item Keyword: data[``keyword'']
    \item Label: data[``label'']
    \item Begin Output:
\end{enumerate}
\end{tcolorbox}

\begin{tcolorbox}[
  title=Generate triples prompt,
  boxrule=0.6pt,
  arc=2pt,
  left=2pt,right=2pt,top=2pt,bottom=2pt,
  fonttitle=\bfseries
]

\textbf{System prompt:}
I will provide a keyword and a section of text content retrieved from fashion-related sources. Your task is to extract structured Knowledge Graph triples based on this content.

Requirements:
\begin{enumerate}
\setlength{\itemsep}{0pt}\setlength{\parsep}{0pt}\setlength{\topsep}{0pt}
    \item Extract triples in the format (head, relation, tail).
    \item  You must ONLY use relations from the following allowed list: [designs, creativeDirectorOf, isCommonlyMadeOf, hasSilhouetteCommonly, hasStyle, influencedBy, originatedIn, usesTechnique, followsPrinciple, hasPart, isAssociatedWith, isCharacterizedBy, isPopularAt, happenIn, pairWith, inspiredBy].
    \item  Adhere to strict type constraints:\\
       - ``originatedIn'': head must be Brand/Person, tail must be Location/Time.\\
       - ``hasPart'': head must be Product/Brand, tail must be Feature/Sub-brand.\\
       - ``isCommonlyMadeOf'': tail must be Material/Fabric.
    \item  Ensure both ``head'' and ``tail'' are specific entities found in or inferred from the text.
    5. Output as a JSON list in the format [{``head'': ``X'', ``relation'': ``Y'', ``tail'': ``Z''}].
    \item Output only the JSON result directly without explanation. If no valid triples are found, output an empty list [].
\end{enumerate}

\textbf{User prompt:}
\begin{enumerate}
\setlength{\itemsep}{0pt}\setlength{\parsep}{0pt}\setlength{\topsep}{0pt}
    \item Keyword: data[``keyword'']
    \item Text Content: data[``combined text'']
    \item Begin Output:
\end{enumerate}
\end{tcolorbox}

\begin{tcolorbox}[
  title=Nodes labeling prompt,
  boxrule=0.6pt,
  arc=2pt,
  left=2pt,right=2pt,top=2pt,bottom=2pt,
  fonttitle=\bfseries
]

\textbf{System prompt:}
I will provide a specific entity name that is missing a category label in the Knowledge Graph, along with a list of valid domain labels. Your task is to categorize the entity using ONLY the provided labels.

Requirements:
\begin{enumerate}
\setlength{\itemsep}{0pt}\setlength{\parsep}{0pt}\setlength{\topsep}{0pt}
\item Select the most specific and accurate label from the ``Available Labels'' list provided in the input.
\item Do NOT invent new labels. You must strictly match one of the strings from the provided list.
\item If the entity is ambiguous, select the most relevant fashion/design-related label.
\item Output as a JSON object in the format: {``name'': ``entity name'', ``label'': ``SelectedLabel''}.
\item Return ONLY the JSON object. Do not include markdown formatting (like ```json), explanations, or notes.
\end{enumerate}

\textbf{User prompt:}
\begin{enumerate}
\setlength{\itemsep}{0pt}\setlength{\parsep}{0pt}\setlength{\topsep}{0pt}
    \item Entity: data[``node name'']
    \item Available Labels: data[``labels str'']
    \item Begin Output:
\end{enumerate}

\end{tcolorbox}
\end{table}

\subsection{FashionEcoKG Initialization}
The initialization of FashionEcoKG systematically extracts domain entities $\mathcal{E}$ and relations $\mathcal{R}$ from a corpus of authoritative fashion textbooks $\mathcal{T}$. To preserve the authority of the source material, which draws on interviews with students at leading Chinese fashion universities, the original metadata is maintained in Chinese; all triples are subsequently translated into English to support broader generalization. Although translation may introduce subtle nuances, modern LLMs substantially reduce cross-lingual barriers, rendering this concern secondary to our primary objectives. Given the advanced capability of LLMs in perceiving textual semantics and resolving well-defined extraction tasks from unstructured text~\cite{gao2025large,feng2025llmkg+}, we adopt an LLM agent-based approach for both Named Entity Recognition (NER) and Relation Detection (RD). A decoupled, sequential pipeline processes the raw textbook content through these two phases, and both are executed chapter-by-chapter to preserve local semantic focus.

For effective NER, the agent is first instructed with an \textbf{expert-crafted ontology} of the fashion ecosystem, which defines eight primary categories and multiple secondary sub-categories, as shown in~\autoref{fig:ontology}. This ontology was co-developed by two fashion experts, audited by a third, and iteratively refined according to the error patterns surfaced during preliminary NER/RD testing. To ensure extraction quality, we benchmarked multiple LLM backbones against manual annotations produced by a fashion-major researcher for a representative chapter, which served as the ground-truth baseline. The model achieving the highest alignment with both the hierarchical ontology and the human-annotated results was then selected as the primary agent for ontologically grounded NER across the textbook corpus.

For the RD phase, the agent is supplied with the raw chapter text, the entity set identified by NER, and a dictionary of \textbf{pre-defined relations} established by domain experts (Supplementary Materials). This closed relational vocabulary constrains the model to detecting edges within a fixed semantic set. Similar to NER, several LLM backbones are compared, and an additional LLM serves as a reflection agent to further refine the outputs. The final production model was selected through human sampling evaluation, ensuring that the resulting triples preserve the structural rigor required for expert-level reasoning. Following NER and RD, the preliminary triples of FashionEcoKG were generated and then subjected to \textbf{manual review}, a labor-intensive validation process totaling approximately 100 man-hours contributed by eight fashion-expert annotators, with a domain-specialized annotator (fashion major) verifying and correcting the generated triples to guarantee data fidelity. Specific prompts applied for NER and RD are given in Table~\ref{tab:KG_initialization_prompt}. 

While the initialization phase ensures that the foundation of the FashionEcoKG is anchored in academically rigorous and high-precision domain knowledge, the resulting graph topology is limited by \textbf{structural sparsity}. Analysis of the initial graph indicates that the condensed nature of textbook corpora often leads to concepts being introduced in isolation, resulting in a fragmented architecture with restricted relational connections. To mitigate these structural bottlenecks and facilitate more complex reasoning, we introduce two additional augmentation phases to bridge gaps between isolated expert-level entities as well as to enrich the graph with stronger relational connections.

\subsection{FashionEcoKG Augmentation}
We propose a two-faceted augmentation stage for our FashionEcoKG with two processes: Cross-domain Relational Augmentation and Generative KG Expansion. 

\textbf{Cross-domain Relational Augmentation.} To address the topological fragmentation identified in the initial graph, we implement a Cross-domain Relational Augmentation phase utilizing a domain-restricted slice of DBpedia. Given the extreme scale and inherent noise of the full DBpedia corpus, we isolate the Fashion category subgraph\footnote{https://dbpedia.org/page/Fashion} as a high-fidelity reference set to maintain efficiency and precision. The augmentation begins with Entity Alignment, where a coarse matching process is executed between the entities distilled in the previous stage and the resources within the DBpedia fashion slice. For each successfully aligned entity, we perform Neighborhood Subgraph Extraction to retrieve all associated triples. To ensure structural integrity, the resulting candidate relations are manually evaluated. Relations that are low quality or semantically divergent from the fashion ecosystem are removed.
The remaining qualified triples are merged into FashionEcoKG, serving as the `structural glue' that converts isolated clusters into a connected, multi-hop topology.

\begin{figure*}[t]
    \centering
    \includegraphics[width=1\linewidth]{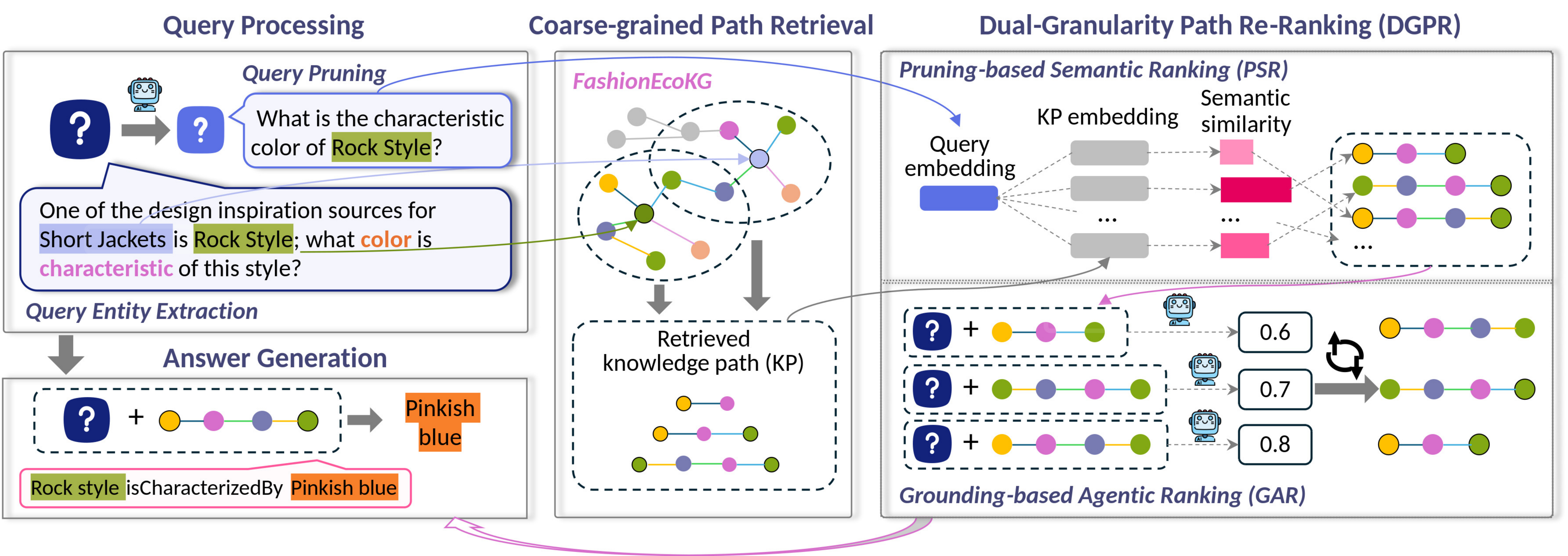}
    \caption{Overview of the PG-RAG Pipeline. Based on coarse-grained path retrieval, PG-RAG introduces a Dual-Granularity Re-ranking module to address conceptual density and query distraction in fashion QA. 
    The PSR module extracts query skeletons for high-recall path retrieval, followed by the GAR module for point-wise candidate path scrutiny against the full query to ensure global relevance.}
    \label{fig:model}
\end{figure*}

\textbf{Generative KG Expansion}
To grow the narrow core into a deep, comprehensive resource, we deploy an LLM-driven expansion pipeline that uses the verified entities of the grounded FashionEcoKG base as seeds for querying open web data. For each seed entity $s$, an automatic search aggregates a corpus of unstructured documents from the open web, which then serve as the dynamic context for an LLM agent. To keep the generated knowledge domain-aligned, we condition the LLM on the relation set established in previous phases. 

Candidate triples undergo two-facet verification before integration. First, the agent checks whether each triple's entities map to categories of the expert-crafted ontology, a domain filter that confines new concepts to the established fashion ecosystem. Second, to prevent graph inflation and preserve topological efficiency, each candidate triple is compared against the current graph state and discarded if redundant. Through retrieval, regulated generation, and verification, FashionEcoKG grows substantially in scale and relational connectivity, enabling expert-level reasoning while preserving authoritative precision. Specific prompts applied for FashionEcoKG expansion are given in Table~\ref{tab:KG_expansion_prompt}. Statistics of the different stages of FashionEcoKG are given in Table~\ref{tab:kg_stac}. 

Beyond the consistent gains FashionEcoKG yields across multiple RAG strategies, we further assess triple-level quality via a multi-agent LLM check: Gemini 2.5 and GPT-5 independently cross-validate the extracted triples, achieving over 94\% individual correctness and 72\% joint agreement. This confirms both the reliability of our automated extraction pipeline and the overall quality of the curated knowledge.

Finally, FashionEcoKG supports incremental updates: newly acquired triples are mapped to established ontological anchors and verified using the same procedure introduced in the augmentation process.

\begin{table}[htbp]
\centering
\caption{FashionEcoKG Statistics}
\label{tab:kg_stac}
% \scalebox{1.0}{
\begin{tabular}{c c c c}
\toprule
&\textbf{KG-V1} &\textbf{KG-V2} &\textbf{KG-V3} \\
\midrule
\# Entity &4930 &5320 &7299 \\
\# Triple &8041 &8770 &10592 \\
\bottomrule
\end{tabular}
% }
\end{table}
\section{Model: Pruning-Grounding Knowledge Graph Retrieval-Augmented Generation}
% We develop a FashionKG-RAG pipeline to bridge the gap between the static parametric memory of LLMs and the structured, domain-specific intelligence of our FashionEcoKG. 
In this section, we introduce our Pruning-Grounding Knowledge Graph Retrieval-Augmented Generation (PG-RAG) method, which leverages the constructed FashionEcoKG for knowledge-intensive fashion QA. 
Given a question $q$, the ultimate task is to design a function $f$ to predict the answer $a$, i.e., $a=f(q)$
We denote our FashionEcoKG as $\mathcal{G}=\{<e_h, r, e_t> | e_h, e_t \in \mathcal{E}, r \in \mathcal{R}\}$, where $\mathcal{E}$ and $\mathcal{R}$ are the set of entities and relations respectively, and $<e_h, r, e_t>$ denotes a triple. 
With FashionEcoKG, we aim to extract effective information $\mathcal{I}$ based on the question with an effective retriever ($g$): $\mathcal{I} = g(q, \mathcal{G})$, thereby helping the generator ($f$) for answer prediction: $a = f(q, \mathcal{I})$.

\subsection{Coarse-grained Path Retrieval}
% \yx{Path-based KG-RAG Pipeline}}
% Given a question $q$, KG-RAG relies a retriever to search for relevant information from a well-constructed KG $\mathcal{K}$. 
Given a well-constructed KG $\mathcal{G}$, our PG-RAG is a path-based RAG pipeline that first encodes the KG into multi-hop paths with maximum path length being $T$, which can be conducted offline. The whole path set can be denoted as $\mathcal{P} = \{\mathcal{P}^{1}, \ldots, \mathcal{P}^T\}$, where $\mathcal{P}^{i}$ denotes the collection of paths with $i$-hop connections. Each path can be indicated with entities in it, which means every entity in the KG is associated with a set of paths, denoted as $\mathcal{P}_e$. 
For example, a 2-hop path can be denoted as $p=e_0 \xrightarrow{r_0}e_1\xrightarrow{r_1}e_2$, $p \in \mathcal{P}_{e_0}$ and $p \in \mathcal{P}_{e_2}$. 
We treat each question $q$ as the original query for the retrieval system. 
To support entity retrieval, we first conduct entity extraction with an LLM agent $f_{ee}$ for a given query to obtain entity mention set $\mathcal{E}_{q} = f_{ee}(q) = \{e_q\}, e_q \in q$. After that, we can retrieve relevant knowledge paths for query $q$ through entity matching: $\mathcal{P}_q=\{p_e | e \in \mathcal{E}_q\}$. For each query $q$, the entity-matching path retrieval results offer coarse-grained relevant candidates. 

\subsection{Dual-Granularity Path Re-Ranking (DGPR)}
The initial retrieval $\mathcal{P}_q$ often suffers from noise due to the conceptual density and descriptive distractors present in fashion queries, as shown in the example illustrated in ~\autoref{fig:model}. Purely entity-based retrieval also neglects relational information. 
To refine this issue, we propose Dual-granularity Path Re-Ranking (DGPR), comprising a semantic-alignment pruning stage and a structural-verification grounding stage.
% To refine this issue, we propose Dual-granularity Path Re-Ranking (DGPR), comprising a pruning stage for semantic alignment and a grounding stage for structural verification.

\subsubsection{Pruning-based Semantic Ranking (PSR)}
Fashion queries typically involve combinational subjects where only a subset of mentions is critical for the reasoning task. To isolate the core intent, we utilize an LLM agent $f_{qp}$ as the \textit{query pruner}, transforming the original query $q$ into a compact query skeleton $q_s = f_{qp}(q)$. This process prunes linguistic noise and non-essential entity mentions that might lead to semantic drift during retrieval. The semantic similarity between the query skeleton $q_s$ and each candidate path $p \in \mathcal{P}_q$ obtained from initial stage is further computed: 
\begin{equation}
    s(q_s, p) = \text{sim}(h_{q_s}, h_p), 
\end{equation}
where $h_{q_s}$ and $h_p$ are semantic embeddings for them respectively, which can be obtained with a pre-trained language model. We select the top-$k_1$ paths $\mathcal{P}_{psr} \subseteq \mathcal{P}_q$ with the highest similarity scores. By prioritizing the skeleton, PSR ensures a high-recall set of path candidates while effectively neutralizing combinational distractors. 

\subsubsection{Grounding-based Agentic Ranking (GAR)}
While PSR identifies semantically related paths, purely embedding-based retrieval often fails to capture the structural logic or overlooks fine-grained constraints. To overcome this limitation, we introduce Grounding-based Agentic Ranking (GAR). This stage utilizes a discriminative evaluator agent $f_{e}$ to perform point-wise scoring on the pruned candidate set $\mathcal{P}_{psr}$ using the original query $q$. The original query is chosen here rather than the simplified skeleton used in PSR to ensure that no expert-level constraints or nuanced requirements are lost during the grounding process. While PSR prioritizes retrieval breadth by neutralizing noise, GAR provides the necessary precision by scrutinizing each candidate path $p_i \in \mathcal{P}_{psr}$ against every detail of the user's intent. Unlike semantic ranking, which relies on global vector proximity, the agent explicitly interprets the relational transitions within a path (e.g., $e_0 \xrightarrow{r_0} e_1 \xrightarrow{r_1} e_2$) to determine if the knowledge chain logically satisfies the query. For each path, the agent assigns a grounding score $\omega_i \in [0, 1]$:
\begin{equation}
    \omega_i = f_{e}(q, p_i), \quad p_i \in \mathcal{P}_{psr}.
\end{equation}
% This stage transforms the retrieval process from a ``similarity match'' into a ``logical verification'' task. 
The final set of retrieved information $\mathcal{I}$ is constructed by selecting the top-$k_2$ paths with the highest evaluation scores:
\begin{equation}
    \mathcal{I} = \{p_i \mid \text{top-}k_2 \text{ sorted by } \omega_i\}.
\end{equation}
By transitioning from the broad semantic filtering of PSR to the rigorous agentic grounding of GAR, the PG-RAG framework ensures that the final generation is anchored in logically verified, highly relevant knowledge.

\begin{table*}[h!]
\centering
\caption{Fashion QA accuracy of compared methods with regard to different question types and complexities. Defi.: Definition, Infl.: Influence, Part.: Partonomy, Char.: Characteristic, Exem.: Exemplification, Insp.: Inspiration, Comp.: Comparison. }
\label{tab:overall_performance}
\renewcommand{\arraystretch}{1}
\resizebox{\linewidth}{!}{
\begin{tabular}{c
    c c c c c c c c
    c c c
    c }
    \toprule
    \multirow{2.5}{*}{\textbf{Model}} & \multicolumn{8}{c}{\textbf{Question Category}} & \multicolumn{3}{c}{\textbf{Complexity}} & \multirow{2}{*}{\textbf{Overall}} \\
    \cmidrule(lr){2-9} \cmidrule(lr){10-12}
    & \textbf{Defi.} & \textbf{Infl.} & \textbf{Part.} & \textbf{Char.} & \textbf{Origin} & \textbf{Exem.} & \textbf{Insp.} & \textbf{Comp.} & \textbf{Easy} & \textbf{Medium} & \textbf{Hard} & \\
    \midrule
    \multicolumn{13}{c}
    {\cellcolor{gray!20}\textit{\textbf{open-source VLMs}}} \\
    \textbf{Qwen2.5-VL-7B ~\cite{qwen2.5}} & 82.32 & 64.10 & 55.14 & 56.27 & 53.94 & 52.73 & 43.82 & 41.38 & 82.01 & 53.75 & 57.76 & 59.86 \\
    \textbf{Qwen3-VL-8B~\cite{qwen3}} & \bestOS{93.91} & 71.79 & 64.49 & 65.37 & 66.76 & 57.58 & 53.93 & 31.03 & \bestOS{93.81} & 65.63 & 59.03 & 69.44 \\
    \textbf{InternVL3-8B~\cite{zhu2025internvl3}} & 86.96 & 60.26 & 62.62 & 64.33 & 61.81 & 47.88 & 44.94 & 31.03 & 87.02 & 60.97 & 56.74 & 64.90 \\
    \textbf{InternVL3.5-8B~\cite{wang2025internvl3_5}} & 92.17 & 66.67 & 58.88 & 63.43 & 65.31 & 56.97 & 55.06 & 37.93 & 92.04 & 63.80 & 57.51 & 67.69 \\ % New Model 1
    \textbf{Gemma-3-12B~\cite{Gemma2025Gemma}} & 88.99 & 62.82 & 65.42 & 66.72 & 58.60 & 58.79 & 50.56 & \bestOS{44.83} & 89.09 & 63.89 & 58.02 & 67.31 \\ % New Model 2
    \textbf{MiniCPM-V-4.5-8B~\cite{yu2025minicpm}} & 92.17 & \bestOS{73.08} & 63.55 & 62.84 & \bestOS{67.35} & 53.94 & 55.06 & 27.59 & 92.04 & 63.35 & \bestOS{60.05} & 67.96 \\
    \rowcolor{blue!15} \textbf{GLM-4.1V-9B~\cite{hong2025glm}} & 92.75 & 70.51 & \bestOS{66.36} & \bestOS{67.46} & 66.76 & \bestOS{61.21} & \bestOS{58.43} & 41.38 & 92.63 & \bestOS{67.82} & \bestOS{60.05} & \bestOS{70.76} \\
    \midrule
    \multicolumn{13}{c}{\cellcolor{gray!20}\textit{\textbf{open-source LLMs}}} \\
    \textbf{DeepSeek-R1-Qwen-7B~\cite{DeepSeekAI2025DeepSeekR1IR}} & 71.01 & 39.74 & 38.32 & 38.21 & 37.32 & 36.36 & 37.08 & 17.24 & 70.50 & 37.29 & 38.68 & 43.76 \\
    \textbf{GLM-4-9B~\cite{Zeng2024ChatGLMAF}} & 75.94 & 62.82 & 42.99 & 50.60 & 58.89 & 47.27 & 34.83 & 27.59 & 75.81 & 51.55 & 49.36 & 55.59 \\
    \textbf{Llama3.1-8B~\cite{dubey2024llama}} & 85.17 & 61.54 & 52.34 & 49.25 & 56.25 & 47.27 & 48.31 & 34.48 & 84.91 & 50.61 & 53.18 & 57.60 \\
    \rowcolor{red!15}\textbf{Qwen2.5-7B~\cite{qwen2.5}} & 93.60 & 70.51 & 57.01 & 62.99 & 63.12 & 54.55 & 51.69 & 34.48 & 93.79 & 61.90 & 58.02 & 67.04 \\
    \rowcolor{yellow!15}\textbf{Qwen3-8B~\cite{qwen3}} & 91.30 & 69.23 & 59.81 & 64.93 & 60.64 & 57.58 & 52.81 & 34.48 & 91.15 & 63.16 & 58.02 & 67.25 \\
    \midrule
    \multicolumn{13}{c}            {\cellcolor{gray!20}\textit{\textbf{closed-source LLMs}}} \\
    \rowcolor{green!15}\textbf{Gemini-2.5 Flash~\cite{comanici2025gemini}} & 92.46 & 74.36 & 64.49 & \bestComm{69.85} & 65.89 & 57.23 & \bestComm{56.18} & 37.93 & 92.63 & 67.49 & 61.83 & 70.94 \\
    \textbf{Grok-4 Fast~\cite{xaiGrok}} & \bestComm{95.65} & 71.79 & \bestComm{75.70} & 67.76 & \bestComm{68.51} & \bestComm{63.03} & 52.81 & \bestComm{48.28} & \bestComm{95.87} & \bestComm{68.28} & \bestComm{63.36} & \bestComm{72.34} \\
    \textbf{GPT-5 Nano~\cite{OpenAIGPT5}} & 93.91 & \bestComm{78.21} & 66.36 & 67.31 & 64.43 & 58.18 & \bestComm{56.18} & 34.48 & 94.10 & 66.64 & 60.05 & 70.32 \\
    % \midrule
    % \textbf{Average} & \textbf{88.24} & \textbf{66.77} & \textbf{59.17} & \textbf{60.55} & \textbf{60.91} & \textbf{53.45} & \textbf{49.70} & \textbf{33.95} & \textbf{88.17} & \textbf{59.89} & \textbf{56.63} & \textbf{64.45} \\
    % \midrule
    % Human & 99.66 & 50.00 & 97.62 & 58.88 & 70.59 & 69.23 & 66.67 & 37.50 & 98.15 & 60.81 & 68.25 & 74.95 \\
    \midrule
    \multicolumn{13}{c}{\cellcolor{gray!20}\textit{\textbf{KG-RAG Methods}}} \\
    % \textbf{LightRAG\textsubscript{Qwen2.5-7B}~\cite{guo2025lightrag}} & 88.53	&72.48	&52.11	&58.88	&63.61	&50.44	&52.94	&53.12	&88.50	&60.86	&55.28	&63.87\\
    % \textbf{LinearRAG\textsubscript{Qwen2.5-7B}~\cite{zhuang2025linearrag}} & 89.33	&62.39	&57.75	&60.48	&59.03	&53.98	&57.98	&43.75	&89.09	&60.47	&56.81	&64.10\\
    \textbf{TOG\textsubscript{Qwen2.5-7B}~\cite{sun2023think}} & 88.24 & 70.91 & 63.38 & 66.26 & 59.85 & 60.00 & 58.47 & 40.62 & 87.57 & 65.78 & 58.73 &67.51 \\
    % \textbf{KAPING\textsubscript{Qwen2.5-7B}~\cite{baek2023knowledge}} & 90.40 & 67.89 & 57.04 & 62.75 & 70.99 & 65.04 & 58.82 & 40.62 & 90.27 & 66.15 & 60.84 & 68.67\\
    \textbf{KAPING\textsubscript{Qwen2.5-7B}~\cite{baek2023knowledge}} & 90.67 & 77.98 & 65.49 & 65.42 & 65.90 & 59.29 & 63.87 & 46.88 & 90.27 & 67.94 & 60.08 & 69.56\\
    \textbf{KG-Triple-RAG\textsubscript{Qwen2.5-7B}} & 89.07 & 73.39 & 63.38 & 66.76 & 70.23 & 60.62 & 68.91 & 46.88 & 88.79 & 70.97 & 57.77 & 70.58  \\
    \textbf{KG-Path-RAG\textsubscript{Qwen2.5-7B}} & 88.80 & 73.39 & 60.56 & 66.62 & 70.99 & 64.60 & 64.71 & 46.88 & 88.20 & 69.26 & 62.57 & 70.63\\
    % Triple-based SS Retrieval & 4.13& 24.20& 6.53& 16.74& 88.79& 70.97& 57.77& 70.58  \\
    % Path-based SS Retrieval & 3.54& 24.82& 16.12& 19.35& 88.20& 69.26& 62.57& 70.63\\
    \midrule
    \multicolumn{13}{c}{\cellcolor{gray!20}\textit{\textbf{PG-RAG Methods}}} \\
    \rowcolor{red!15} \textbf{PG-RAG}\textsubscript{Qwen2.5-7B}&92.53& 83.49& 69.01& 68.36& 73.28& 67.26& 73.95& 46.88& 92.04& 73.23& 64.88& 74.17\\
    \rowcolor{yellow!15} \textbf{PG-RAG}\textsubscript{Qwen3-8B}&89.87& \textbf{84.40}& 71.13& 70.36& 70.48& 69.91& 73.95& \textbf{56.25}& 89.09& 72.92& 68.91& 74.50\\
    \rowcolor{blue!15} \textbf{PG-RAG}\textsubscript{GLM-4.1V-9B}&90.67& 81.65& 74.65& 73.70& 73.79& 73.01& \textbf{78.15}& \textbf{56.25}& 89.97& 75.64& 72.17& 77.06\\
    \rowcolor{green!15} \textbf{PG-RAG}\textsubscript{Gemini-2.5 Flash} & \textbf{94.13}& 83.33& \textbf{83.10}& \textbf{78.50}& \textbf{75.83}& \textbf{78.32}& 76.47& 50.00& \textbf{93.81}& \textbf{80.47}& \textbf{72.88}& \textbf{80.74}\\
    \bottomrule
\end{tabular}
}
\end{table*}
% \subsection{Research Questions (RQs)}

\begin{table*}[t]
\caption{PG-RAG prompts}
\label{tab:prompts_rag}
\centering
\scriptsize

\begin{tcolorbox}
[
  % title=PG-RAG Prompt,
  boxrule=0.6pt,
  arc=2pt,
  left=2pt,right=2pt,top=2pt,bottom=2pt,
  fonttitle=\bfseries
]

\begin{minipage}[t]{0.48\linewidth}
\tcbsubtitle{Query pruning prompt}
\textbf{System prompt:}
Your task is to perform simplification on the question and extract the main question that best simplifies the original question. If the question is already a simple question, just use the original question as a simplified question.

Format: Use Markdown headings (``\# 1. Analysis'' and ``\# 2. Final Answer'') to structure your output clearly.

\textbf{CRITICAL}: Conclude your analysis by providing the single best simplified question after ``\# 2. Final Answer''.

\textbf{User prompt:}
Question: data["question"]. Begin Output:
\end{minipage}
\hfill
\begin{minipage}[t]{0.48\linewidth}
\tcbsubtitle{Topic entity extraction prompt}
\textbf{System prompt:}
You are an expert in text entity recognition. Extract all nouns from the given question. Output in JSON format, e.g., \{``entities'': [``entity1'', ``entity2'']\}.

Note: 1. Each line must be valid JSON. 2. If no entities are found, output ``No entities found''.

\textbf{User prompt:}
Question: data[``question'']. Begin Output:
\end{minipage}

\vspace{2pt}

\begin{minipage}[t]{0.48\linewidth}
\tcbsubtitle{No RAG QA prompt}
 \textbf{System prompt:}
     You are an expert fashion connoisseur. Your task is to analyze the question and context, and then strictly follow the steps below to output your final answer:
     \begin{enumerate}
\setlength{\itemsep}{0pt}\setlength{\parsep}{0pt}\setlength{\topsep}{0pt}
         \item Internal Knowledge and Analysis: First, reason step-by-step why each option is correct or incorrect based on both the provided context and your internal knowledge
         \item Selection:Based on your analysis, select the single best option. If, after thorough reasoning, you cannot confidently choose one, randomly select one option (A, B, C, or D) as your final answer to avoid non-response.
         \item Format: Use Markdown headings (1. Analysis and 2. Final Answer) to structure your output clearly.
         \item Final Answer: Conclude your analysis by providing the single best option letter inside $<Answer>$ and $<Answer>$ tags.
     \end{enumerate}

     \textbf{CRITICAL:} The entire output MUST strictly start with the $<Answer>$ tag, followed by a newline, then your detailed analysis.

\textbf{User prompt:}
\begin{enumerate}
\setlength{\itemsep}{0pt}\setlength{\parsep}{0pt}\setlength{\topsep}{0pt}
    \item Question: {data[``question'']}
    \item Options: {data[``options'']}
    \item Begin Output:
\end{enumerate}
\end{minipage}
\hfill
\begin{minipage}[t]{0.48\linewidth}
\tcbsubtitle{RAG QA prompt}
    \textbf{System prompt:}
     You are an expert fashion connoisseur. Your task is to analyze the question and context, and then strictly follow the steps below to output your final answer:
     \begin{enumerate}[itemsep=0pt, topsep=0pt, parsep=0pt]
    \item Provided Information: The context includes several contexts retrieved by the question. The contexts are provided in decreased similarity with the question.
    \item Knowledge and Analysis: First, reason step-by-step why each option is correct or incorrect based on **both the provided context and your internal knowledge.
    \item Selection: Based on your analysis, select the single best option. If, after thorough reasoning, you cannot confidently choose one, randomly select one option (A, B, C, or D) as your final answer to avoid non-response.
    \item Format: Use Markdown headings (1. Analysis and 2. Final Answer) to structure your output clearly.
    \item Final Answer: Conclude your analysis by providing the single best option letter inside $<Answer>$ and $<Answer>$ tags.
    \end{enumerate}
    \textbf{CRITICAL:} The entire output MUST strictly start with the $<Answer>$ tag, followed by a newline, then your detailed analysis.  Remember to conclude your analysis by providing the single best option letter inside $<Answer>$ and $<Answer>$ tags.

 \textbf{User prompt:}
 \begin{enumerate}
\setlength{\itemsep}{0pt}\setlength{\parsep}{0pt}\setlength{\topsep}{0pt}
     \item Context: data[``context'']
      \item Question: data[``question'']
      \item Options: data[``options'']
      \item Begin Output:
  \end{enumerate}
\end{minipage}

\end{tcolorbox}
\end{table*}

\section{Experiments}
In this section, we evaluate the effectiveness of our proposed PG-RAG framework through extensive experiments. 

\begin{figure}[t]
    \centering
    \includegraphics[width=1\linewidth]{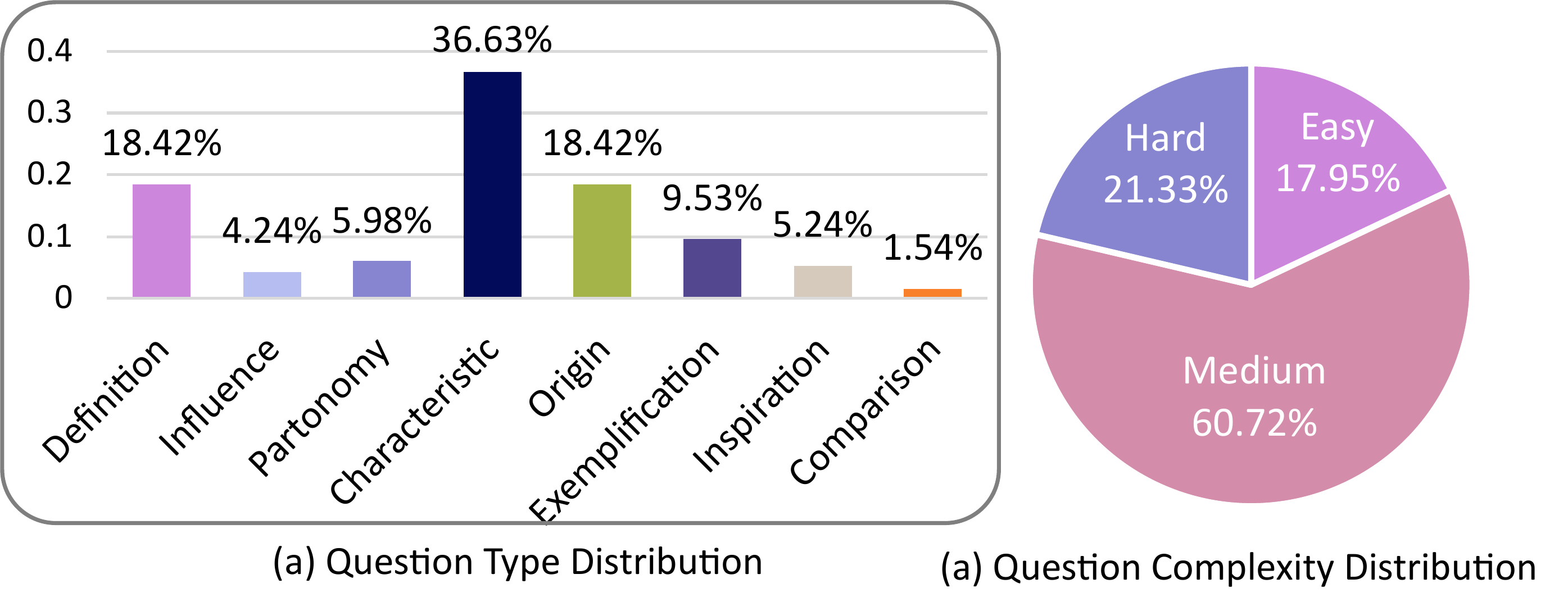}
    \caption{Distribution of question categories (Left) and question complexity (Right) in the tested fashion QA dataset.}
    \label{fig:dataset}
    \vspace{-15pt}
\end{figure}

\subsection{Experimental Setups}
\noindent \textbf{Implementation Details.}
We implement the proposed method using the open-source PyTorch~\cite{paszke2019pytorch} framework.
All experiments are conducted on an NVIDIA RTX 3090 GPU. We adopt Qwen2.5-7B-Instruct~\cite{qwen2.5} as the pruning model, SBERT~\cite{reimers2019sentence}
% and Contriever~\cite{izacard2021unsupervised} 
as the semantic embedding model, and Qwen3-Reranker-4B~\cite{qwen3} as the GAR model.
The top 10 paths retrieved by the PSR stage are retained as candidates for GAR, and the top-ranked GAR result is used as the context for question answering. The candidates for coarse-grained path retrieval include 1-hop, 2-hop, and 3-hop paths. For FashionEcoKG construction, we evaluate three closed-source LLM backbones: Gemini-2.5 Pro, Claude Sonnet 4, and GPT-5. Gemini-2.5 Pro is selected as the primary agent for both NER and RD tasks across all stages. 
% All prompts used to instruct the LLM agents are provided in Supplementary Materials I.
% Appendix~\ref{sec:appdix:prompt}.

\begin{figure}[t]
    \centering
        \includegraphics[width=\linewidth]{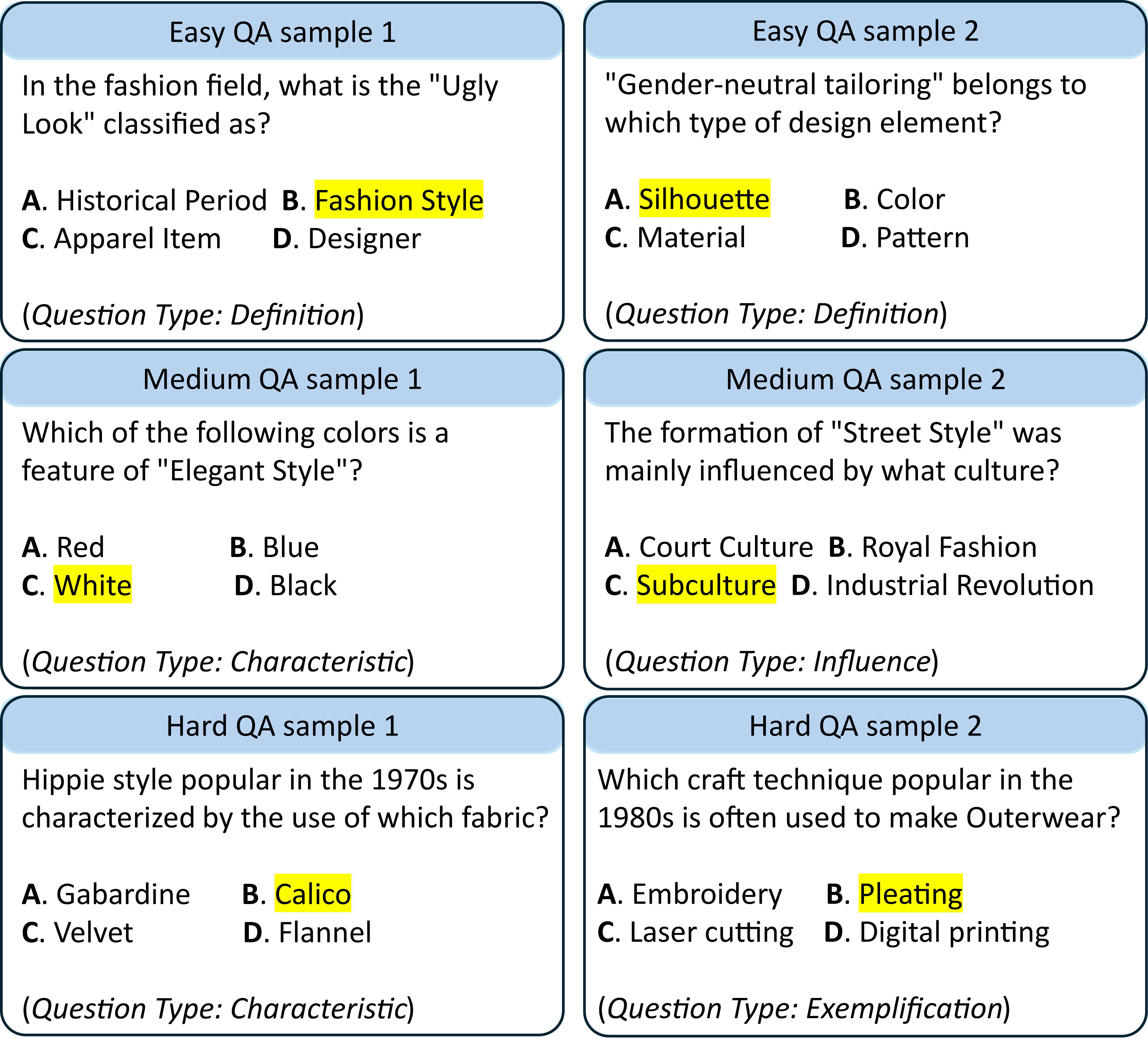}
        \caption{Fashion QA examples with different question types and levels of complexity. The correct answer is highlighted in yellow.}
        \label{fig:datacase}
\end{figure}

\noindent \textbf{Datasets.}
We construct a question-answering dataset using a KG-driven multi-complexity generation pipeline, following a methodology similar to that of concurrent work~\cite{tatarinov2025kg}.
Specifically, we employ LLM agents to generate questions grounded in knowledge paths with different hop-level complexities in the KG.
% Building on the FashionEcoKG, we employed LLM agents to formulate questions grounded in various multi-hop relational paths within the graph. 
The resulting dataset contains 2,153 multiple-choice questions (MCQs) across three complexity levels: Easy (0-hop), Medium (1-hop), and Hard ($\geq$2-hop).
Each sample consists of a ground-truth answer and three plausible distractors. To ensure data integrity, all generated samples are manually verified and refined to correct hallucinations and structural errors, including low-quality distractors, disordered options, answer leakage in the question stem, and logically weak phrasing.
The questions are further categorized into eight types to support fine-grained analysis of model performance across multiple dimensions. In Fig.~\ref{fig:datacase} we illustrate several QA examples with different question types and complexity levels. 
% Appendix~\ref{sec:appdix:qa_samples}.
To complement the synthetic benchmark and to mitigate the concern that questions generated in reverse from a graph may overfit graph structure or retrieval paths, we prepare a tiny human-authored testing subset FashionQA-mini (200 QA samples). This subset is generated independently by human annotators without referring to FashionKG while using only the original textbook as reference material.
% \ly{To further examine whether models can generalize beyond idealized KG-derived question patterns, we additionally construct a human-authored testing subset named FashionQA-mini. More detailed analysis on this is provided in Supplementary Materials F.}

\noindent \textbf{Baselines.}
To evaluate the effectiveness of the proposed PG-RAG model, we compare it with a comprehensive set of baselines, including mainstream Vision-Language Models (VLMs), LLMs, and established KG-RAG methods.
For \textbf{open-source VLMs}, we consider Qwen 2.5VL~\cite{qwen2.5}, Qwen3-VL~\cite{qwen3}, InternVL3~\cite{zhu2025internvl3},InternVL3.5~\cite{wang2025internvl3_5} Gemma-3~\cite{Gemma2025Gemma}, MiniCPM-V-4.5~\cite{yu2025minicpm}, and GLM-4.1V~\cite{hong2025glm}. 
% These models are selected because the fashion domain is vision-intensive, making multi-modal perception a potentially critical factor for addressing complex queries.
Since fashion is a vision-intensive domain, these models are included because their multimodal training may help them acquire and represent fashion knowledge more effectively than text-only LLMs.
For \textbf{open-source LLMs}, we include DeepSeek-R1-Qwen~\cite{DeepSeekAI2025DeepSeekR1IR}, GLM-4~\cite{Zeng2024ChatGLMAF}, LLaMA 3.1~\cite{dubey2024llama}, Qwen 2.5~\cite{qwen2.5}, and Qwen3~\cite{zheng2025empirical,qwen3}.
For \textbf{closed-source LLMs}, we use Gemini-2.5 Flash~\cite{comanici2025gemini, team2023gemini}, Grok-4 Fast~\cite{xaiGrok}, and GPT-5 Nano~\cite{OpenAIGPT5}.

We further include several training-free KG-RAG baselines. In addition to basic Triple-RAG and Path-RAG, which rely on triple- and path-level semantic similarity for retrieval, we compare our method with KAPING~\cite{baek2023knowledge} for entity-based triple retrieval and ToG~\cite{sun2023think} for structured reasoning over graphs.
We exclude approaches that construct their own KGs from scratch, because their performance would be confounded by differences in KG construction quality. To ensure a fair evaluation of retrieval and reasoning, we compare only methods that operate directly on our FashionEcoKG.

% \subsection{Evaluation Metrics}
\noindent \textbf{Evaluation Metrics.}
We evaluate retrieval performance using the Pseudo Relevance Recall (PRRecall@K)~\cite{lin2023fine,luo2021weakly} metric. A document is considered pseudo-relevant if it contains any answers. PRRecall@K measures whether the retrieved $K$ documents include at least one pseudo-relevant document. For QA performance, we report the accuracy. 

\subsection{Fashion QA Performance}
The experimental results summarized in ~\autoref{tab:overall_performance} reveal several key insights into the efficacy of our proposed method. 
First, PG-RAG consistently achieves the best performance across evaluation settings, outperforming both vanilla LLMs and existing KG-RAG baselines. This advantage demonstrates the effectiveness of DGPR in identifying high-quality reasoning paths that contain useful knowledge.
The effectiveness of KG path-based retrieval is further supported by the qualitative cases will be discussed in Sec.~\ref{sec:case}.
% Appendix~\ref{sec:appdix:case}.
\begin{table*}[t]
\centering
\caption{QA accuracy comparison with graph-based RAG baselines. (All methods use Qwen-2.5-7B as the QA model. Best result per column is in \textbf{bold}.}
\label{tab:graph_rag_baselines}
\resizebox{\linewidth}{!}{
\begin{tabular}{lcccccccccccc}
\toprule
\multirow{2.5}{*}{\textbf{Model}} & \multicolumn{8}{c}{\textbf{Question Category}} & \multicolumn{3}{c}{\textbf{Complexity}} & \multirow{2}{*}{\textbf{Overall}} \\
\cmidrule(lr){2-9} \cmidrule(lr){10-12}
& \textbf{Defi.} & \textbf{Infl.} & \textbf{Part.} & \textbf{Char.} & \textbf{Origin} & \textbf{Exem.} & \textbf{Insp.} & \textbf{Comp.} & \textbf{Easy} & \textbf{Medium} & \textbf{Hard} & \\
\midrule
No RAG & \textbf{93.60} & 70.51 & 57.01 & 62.99 & 63.12  & 54.55 & 51.69 & 34.48 & \textbf{93.79} & 61.90 & 58.02 & 67.04 \\
LightRAG~\cite{guo2025lightrag}  & 88.53 & 72.48 & 52.11 & 58.88 & 63.61 & 50.44 & 52.94 & 53.12 & 88.50 & 60.86 & 55.28 & 63.87 \\
LinearRAG~\cite{zhuang2025linearrag} & 89.33 & 62.39 & 57.75 & 60.48 & 59.03 & 53.98 & 57.98 & 43.75 & 89.09 & 60.47 & 56.81 & 64.10 \\
PG-RAG &92.53 & \textbf{83.49} & \textbf{69.01} & \textbf{68.36} & \textbf{73.28} & \textbf{67.26} & \textbf{73.95} & 46.88 & 92.04 & \textbf{73.23} & \textbf{64.88} & \textbf{74.17}\\
\bottomrule
\end{tabular}
}
\end{table*}

When examining the impact of different generators, we observe that the backbone model's inherent capabilities remain influential; GLM-4.1-9B, which is the strongest no-RAG baseline, also achieves the best overall performance when integrated into the PG-RAG pipeline, notably surpassing variants that use Qwen as the generator. 
Furthermore, our comparison of various LLMs and VLMs in no-RAG settings provides a broader understanding of how different architectures handle specialized fashion knowledge. 
We find that performance varies significantly across different versions of the same backbone, with closed-source models generally exhibiting stronger reasoning and domain knowledge than open-source alternatives.

To mitigate overfitting to graph structures in synthetic benchmarks, FashionQA-mini provides 200 human-written questions constructed independently from textbooks without FashionKG access. Four annotators formulate natural, challenging queries testing reasoning, domain expertise, and long-tail concepts, serving as a robustness check for real-world generalization beyond KG-based generation. On FashionQA-mini, our method achieves the best QA accuracy of 85.5\% among all compared settings. It outperforms the No RAG baseline by 13.5\% and KAPING by 6.5\%. Since FashionQA-mini is manually constructed without referring to FashionEcoKG, this improvement suggests that our method generalizes beyond KG-derived question patterns to realistic, human-written fashion-domain questions. 

\textbf{Comparison with Graph-based RAG Baselines}:
We additionally evaluate two representative graph-based RAG methods, LightRAG~\cite{guo2025lightrag} and LinearRAG~\cite{zhuang2025linearrag}. Both methods construct graph indexes to enhance retrieval and have shown effectiveness on relatively well-organized corpora and standard QA benchmarks. However, they are not specifically designed for knowledge-intensive fashion QA, where critical information is embedded in heterogeneous, descriptive, and weakly structured raw text.

As shown in~\autoref{tab:graph_rag_baselines}, both methods underperform the No-RAG base LLM in aggregate accuracy, with scores of 63.87 and 64.10 compared to 67.04. Although they bring improvements on some question types that benefit from relational reasoning, such as Com., and for LinearRAG also Part. and Insp., these gains are outweighed by performance drops on Defi., Char., and Exam. This suggests that graph indexes automatically constructed from weakly structured fashion text can help when explicit relational reasoning is required, but may introduce noise or retrieval mismatches for questions that the base LLM can already answer reliably.

We attribute this degradation to the nature of fashion knowledge. Unlike cleaner corpora where entities and relations can be readily extracted, fashion knowledge is often latent, fine-grained, and context-dependent, appearing within descriptive explanations rather than explicit relational structures. Existing graph-based RAG methods therefore struggle to uncover and exploit such hidden knowledge effectively. These results further motivate our domain-specific knowledge modeling strategy: FashionEcoKG explicitly organizes deeply embedded fashion knowledge into a structured, ecosystem-level representation, while our PG-RAG framework retrieves and reasons over this structure more effectively.

\begin{table*}[h!]
\centering
\caption{
Ablation study results. OSR means PSR with the Original query, while PAR means GAR with the Pruning query. }
\label{tab:module-ablations}
\renewcommand{\arraystretch}{1}
\resizebox{\linewidth}{!}{
\begin{tabular}{
     l c c c c c c c c |c c c c }
    \toprule
     % \multirow{2.5}{*}{\textbf{Ablation}} 
     &   \multicolumn{4}{c}{\textbf{PSRecall@1}} &   \multicolumn{8}{c}{\textbf{Accuracy}} \\
     \cmidrule(lr){2-5} \cmidrule(lr){6-13} 
      &\multicolumn{4}{c}{} &   \multicolumn{4}{c|}{\textbf{Qwen-2.5-7B}} &   \multicolumn{4}{c}{\textbf{Gemini-2.5 Flash}} \\

     \textbf{Ablation} &\textbf{Easy} & \textbf{Medium} & \textbf{Hard} & \textbf{Overall}  &  \textbf{Easy} & \textbf{Medium} & \textbf{Hard} & \textbf{Overall} &  \textbf{Easy} & \textbf{Medium} & \textbf{Hard} & \textbf{Overall}\\
    \midrule
      No RAG & - & -& -&- &90.27& 64.90& 57.97& 67.23&92.63& 67.49&61.83&70.94 \\

    Ours (w/o GAR)  &  2.36& 24.12& 17.66& 19.11& 91.15& 71.28& 62.57& 72.31&93.22& 78.13& 70.25& 78.60\\
    Ours (w OSR) &2.65& 35.25& 25.53& 27.74 &89.38& \textbf{74.32} &62.19 &73.75 & \textbf{94.10}& 80.16& 70.06& 79.91 \\
    Ours (w PAR) &2.36& 34.47&\textbf{27.64}& 27.74& 89.68& 71.36& \textbf{66.22} &73.01& 93.22& 79.77& \textbf{74.28}& 80.56\\
    Ours& \textbf{2.95}& \textbf{35.49}& \textbf{27.64}& \textbf{28.44}& \textbf{92.04}& 73.23 & 64.88& \textbf{74.17}& 93.81 & \textbf{80.47}& 72.88& \textbf{80.74}\\
    % \midrule
    % No RAG & -& -& -& -&92.63& 67.49&61.83&70.94 & 93.81& 80.47& 72.88& 80.74\\
    % Ours w/o GAR & 2.36& 24.12& 17.66& 19.11& 93.22& 78.13& 70.25& 78.60\\
    % Ours w OSR& 2.65&  35.25&  25.53&  27.74& 94.10& 80.16& 70.06& 79.91 \\
    % Ours w PAR& 2.36& 34.47& 27.64& 27.74& 93.22& 79.77& 74.28& 80.56\\
    % Ours& 2.95& 35.49& 27.69& 28.45& 93.81& 80.47& 72.88& 80.74\\
    \bottomrule
\end{tabular}
}
\end{table*}

% \begin{table}[htbp]
% \centering
% \caption{QA Accuracy of graph-based RAG baselines. (All methods use Qwen-2.5-7B as the QA model.)}
% \label{tab:graph_rag_baselines}
% \begin{tabular}{lcccc}
% \toprule
% \textbf{Method} & \textbf{Easy} & \textbf{Medium} & \textbf{Hard} & \textbf{Overall} \\
% \midrule
% No RAG    & \textbf{93.79} & 61.90 & 58.02 & 67.04 \\
% LightRAG  & 88.50 & 60.86 & 55.28 & 63.87 \\
% LinearRAG & 89.09 & 60.47 & 56.81 & 64.10 \\
% PG-RAG    & 92.04 & \textbf{73.23} & \textbf{64.88} & \textbf{74.17} \\
% \bottomrule
% \end{tabular}
% \end{table}

\subsection{Ablation Studies on PG-RAG}
In ~\autoref{tab:module-ablations}, we analyze the impact of different modules of PG-RAG. Specifically, we evaluate the following configurations (using Qwen-2.5 / Gemini-2.5 as the question answering model): 
(1) \textbf{No RAG}: answering without using any retrieved contexts. 
(2) \textbf{Ours (w/o GAR)}: using only the PSR retrieval results. 
(3) \textbf{Ours (w OSR)}: using DGPR retrieval results but removing query pruning in the PSR stage, that is, using the original query to compute semantic similarity with candidate paths. 
(4) \textbf{Ours (w PAR)}: using the pruned query instead of the original one to compute GAR ranking scores. 
(5) \textbf{Ours}: PG-RAG.

The experimental results lead to three main observations.
(1) The two ranking modules, PSR and GAR, both contribute to QA performance. As shown in the second line, PSR alone improves the overall QA performance by 5.08\% (Qwen-2.5) and 7.66\% (Gemini-2.5), indicating the effectiveness of the augmented knowledge.
(2) As shown in the last line, adding the GAR module further brings up the performance by 1.86\% (Qwen-2.5) and 2.14\% (Gemini-2.5). The GAR module significantly improves Top-1 PSRecall by computing fine-grained scores, especially for medium- and hard-level queries. 
(3)  Query pruning is effective in PSR, as reflected by the results of Ours (w OSR), but has a negative effect in GAR. These findings are consistent with our motivation for designing the DGPR module.

\begin{figure}[t]
    \centering
    \includegraphics[width=1\linewidth]{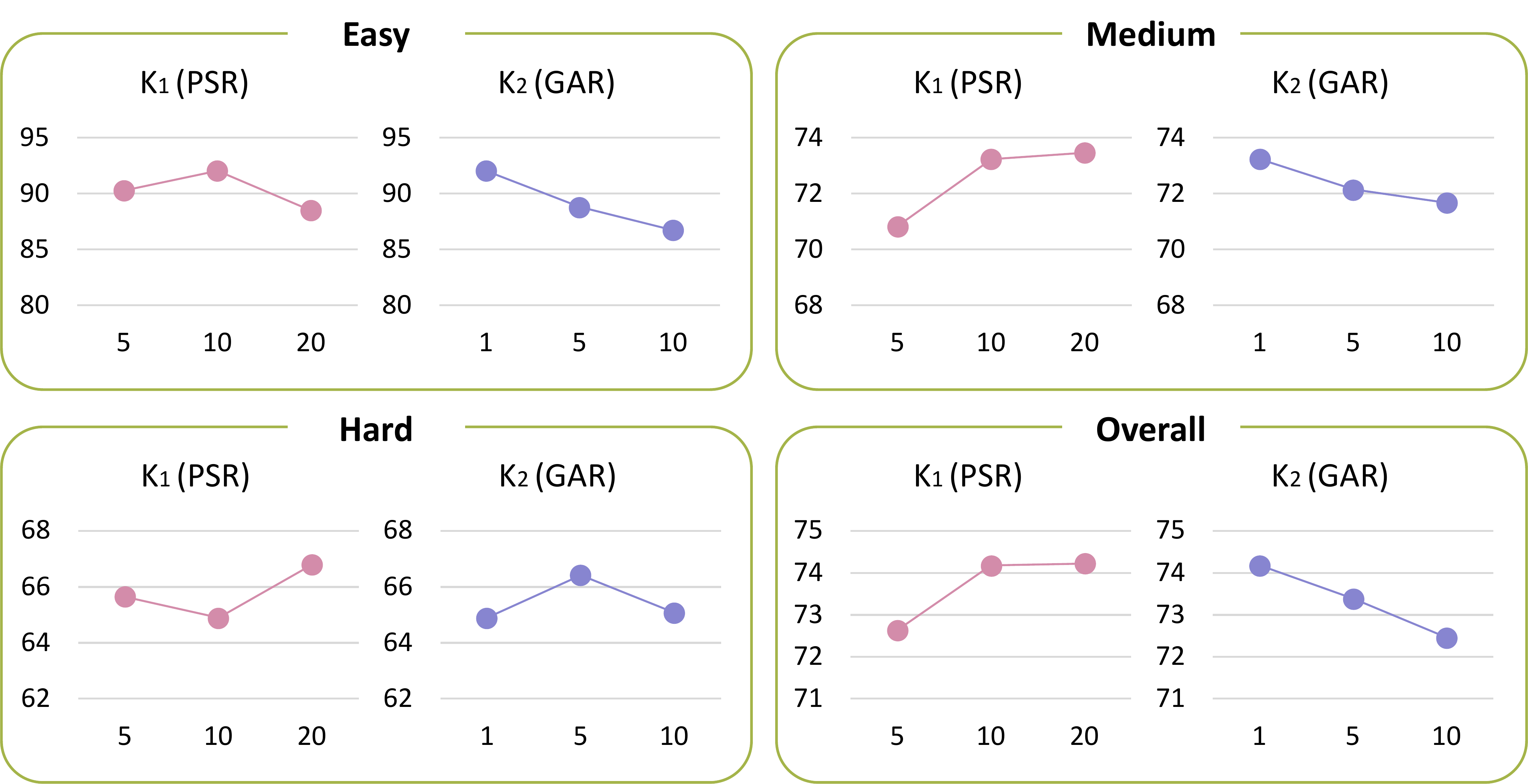}
    \caption{QA accuracy of PG-RAG with different $k_1$, $k_2$ settings.}
    \label{fig:k1k2}
\end{figure}

We further evaluate the effects of the number of paths retained after PSR ($k_1$) and GAR ($k_2$).
As shown in~\autoref{fig:k1k2}, the optimal settings depend on question complexity.
For Easy and Medium cases, the best results are achieved with $k_1=10$ and $k_2=1$, indicating that a single accurately located path is sufficient for simpler queries.
For Hard cases, performance peaks at $k_1=20$ and $k_2=5$, suggesting that complex multi-hop reasoning requires a broader set of knowledge paths to provide sufficient context and satisfy expert-level constraints.

\begin{table}[h!]
\centering
\caption{QA accuracy of RAG methods with different knowledge sources and retriever backbones.}
\label{tab:KG_ablation}
\renewcommand{\arraystretch}{1}
\resizebox{\linewidth}{!}{
\begin{tabular}{c c c c c c c c}
    \toprule
    \textbf{Retriever} &\textbf{K-Type} & \textbf{K-Source} &\textbf{Easy} & \textbf{Medium} & \textbf{Hard} & \textbf{Overall} & \textbf{Impro.} \\
    \midrule
    \multirow{4}{*}{\textbf{SBERT}} &\multirow{2}{*}{\textbf{Chunks}} 
    &Len=100  &88.79 &59.38& 58.35& 63.78 &-5.13\%\\
    &~&Len=1000 & 88.20& 57.98& 54.89& 62.00 &- 7.78\%\\
    \cmidrule(lr){2-8}
    &\multirow{2}{*}{\textbf{KG}} &DBPediaFashion & 87.32& 61.63& 54.13& 63.87 &- 5.00\%\\
    % &~& FashionEcoKG v1 & 4.42& 17.59& 3.45& 12.07& 89.68& 69.42& 56.43& 69.46 &+ 3.32\%\\
    %  &~& FashionEcoKG v2&4.13& 17.59& 3.26& 11.98& 90.27& 69.34& 58.35& 69.98 &+ 4.09\%   \\
    &~ &FashionEcoKG & 88.79& 70.97& 57.77& 70.58 & \cellcolor{red!15} + 4.98\% \\
    \midrule
    \multirow{4}{*}{\textbf{Contriever}} &\multirow{2}{*}{\textbf{Chunks}} 
    &Len=100 & 89.38& 60.54& 54.13 &63.54 &- 5.49\%\\
    &~ &Len=1000 & 85.84& 60.39& 52.98& 62.61&- 6.83\%\\
    \cmidrule(lr){2-8}
    &\multirow{2}{*}{\textbf{KG}} &DBPediaFashion & 89.68& 60.86& 57.97& 64.71&- 3.75\%\\
    % &~ & FashionEcoKG v1 &2.36& 12.45& 2.69& 8.48& 91.15& 66.77& 53.55& 67.41 &+ 0.27\%\\
    %  &~ & FashionEcoKG v2& 2.06& 12.37& 2.88& 8.44& 89.38& 67.47& 57.77& 68.58   &+ 2.01\%\\
    &~ &FashionEcoKG  & 89.68& 67.70& 57.58& 68.72  & \cellcolor{red!15} + 2.22\% \\
    \midrule 
    \multicolumn{2}{c}{\textbf{LightRAG}} 
    % & 88.53 & 72.48 & 52.11 & 58.88 & 63.61 & 50.44 & 52.94 & 53.12 
    & Len=1000 & 88.50 & 60.86 & 55.28 & 63.87  &-4.72\%\\
    \multicolumn{2}{c}{\textbf{LinearRAG}}
    % & 89.33 & 62.39 & 57.75 & 60.48 & 59.03 & 53.98 & 57.98 & 43.75 
    & Len=100 & 89.09 & 60.47 & 56.81 & 64.10 &-4.39\%\\
    % \multicolumn{2}{c}{\textbf{KG2\textsubscript{Qwen2.5-7B}~\cite{zhu2025knowledge}}} & len=100
    % % & 89.07 & 60.55 & 47.89 & 54.87 & 56.49 & 45.58 & 48.74 & 34.38 
    % & 89.68 & 54.71 & 51.06 & 59.35 \\
\bottomrule
\end{tabular}
}
\end{table}

\subsection{Ablation Studies on FashionEcoKG}
To validate the essential role of our knowledge source, we evaluated RAG performance across three distinct formats: our FashionEcoKG, raw Fashion Textbook Chunks, and DBPediaFashion using two retriever backbones. 
To ensure a fair and direct comparison, we use a basic semantic-similarity retrieval strategy across all resources: triple-level retrieval for the KGs and chunk-level retrieval for the text-based resources. 
From the experimental results shown in \autoref{tab:KG_ablation}, we have the following observations: 
(1) Traditional text-based RAG consistently degraded performance (dropping accuracy by 5–7\%). Larger chunks introduced excessive noise, obscuring relevant information.
(2) DBPediaFashion also had a negative impact. Despite its large scale (1,072,977 nodes, 2,691 relations, and over 2.36 million triples), it lacks the professional precision of FashionEcoKG and contains too much noise and redundancy. 
(3) Our FashionEcoKG consistently showed positive results. Its success stems from its knowledge compactness and authoritative, expert-anchored content.

We further compare two advanced KG-RAG methods on textbook chunks, namely LightRAG~\cite{guo2025lightrag} and LinearRAG~\cite{zhuang2025linearrag}, which construct KG based on raw chunks in their distinct manner. 
While these sophisticated methods outperformed naive semantic retrieval, they still failed to achieve a positive net effect on LLM answering accuracy. This observation suggests that in expert-intensive domains characterized by high conceptual density, imprecise knowledge extraction or the introduction of irrelevant structural noise can have a detrimental side effect on the generation process. 
Even advanced graph-based indexing cannot fully overcome the inherent noise and fragmentation in raw text, further underscoring the critical necessity of our structured, high-fidelity FashionEcoKG.

\begin{figure}[t]
    \centering
    \includegraphics[width=1\linewidth]{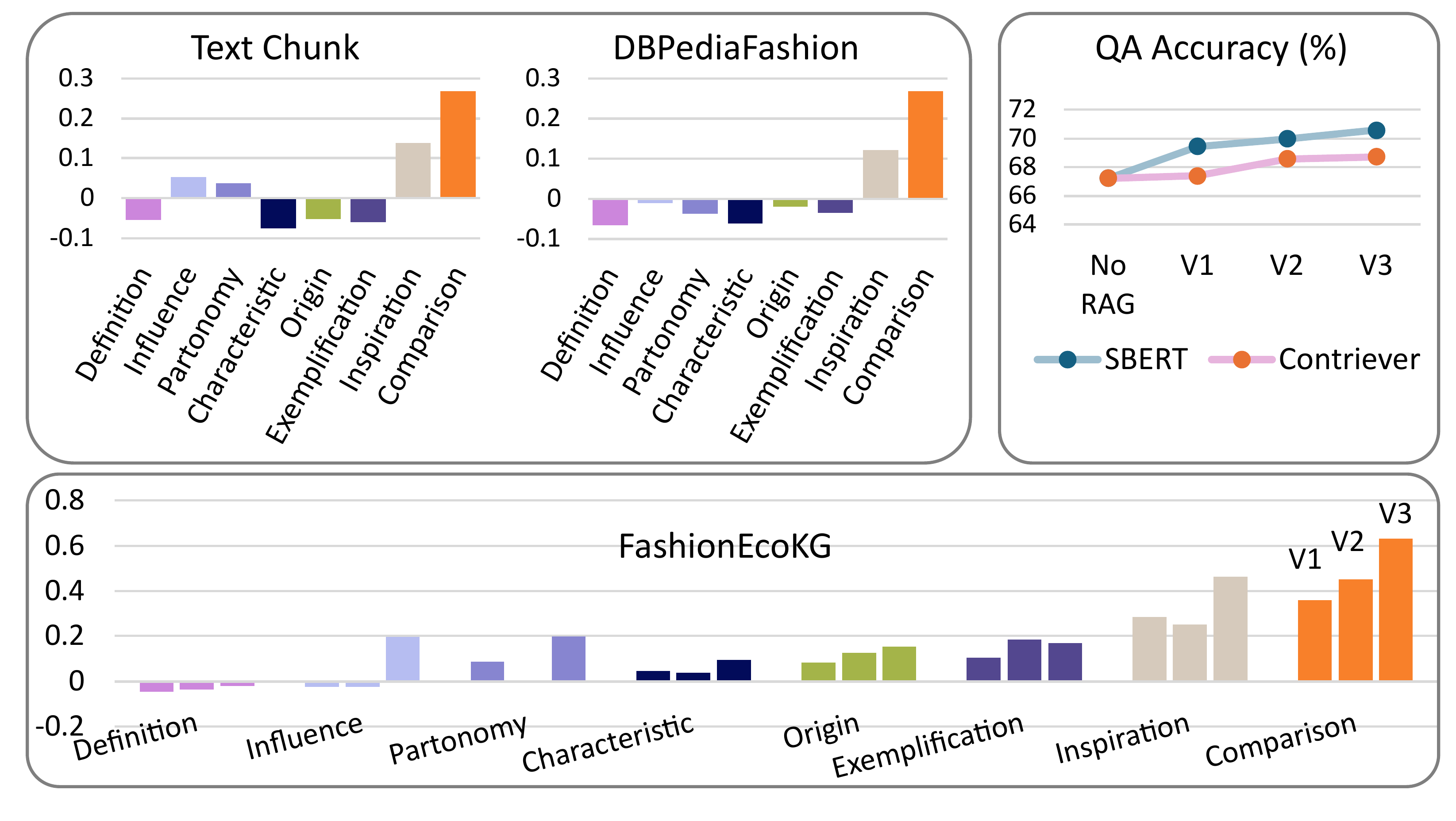}
    \caption{QA accuracy improvement for different question types with different knowledge sources (bars) and overall QA accuracy with three versions of FashionEcoKG under two retriever backbones (chart).}
    \label{fig:different_KG_for_qa_types}
\end{figure}

To investigate the effectiveness of our KG construction pipeline, we experiment on the three sequential versions of FashionEcoKG (V1: Initialized Extraction, V2: Augmentation, V3: Expansion) alongside an analysis across eight question types in \autoref{fig:different_KG_for_qa_types}. 
From the experimental results, we can observe that: 
(1) QA accuracy improved consistently from V1 to V3 across both retrievers. This trend empirically validates our hierarchical construction pipeline, proving that each stage, from foundational expertise to relational densification, directly enhances reasoning capacity.
(2) The final FashionEcoKG (V3) achieved positive gains in 7 out of 8 categories, representing the most robust performance among all sources. In contrast, DBPediaFashion provided only marginal support in \textit{Inspiration} and \textit{Comparison}, while text chunks often caused detrimental side effects in categories such as \textit{Origin} and \textit{Exemplification}.
(3) The highest gains were observed in \textit{Inspiration} and \textit{Comparison}, suggesting that while LLMs possess basic fashion facts, they lack the structural intelligence required for complex cross-concept synthesis. Conversely, no source improved performance in the \textit{Definition} domain, implying that LLMs' parametric memory is already sufficient for basic terminology, rendering external augmentation redundant for simple identification.

\subsection{Theoretical Cost Analysis of PG-RAG}

We conduct efficiency and cost comparisons for different retrieval methods. For PG-RAG, the embeddings of questions, nodes, and knowledge paths are pre-computed offline. Therefore, the online efficiency bottleneck mainly comes from node similarity search and LLM-based reranking/answer generation. Specifically, PG-RAG involves a one-time preprocessing stage and a per-query online inference stage, with notations defined in~\autoref{tab:cost-notations}. 

\begin{table}[htbp]
\centering
\caption{Notations for cost analysis.}
\label{tab:cost-notations}
\small
\begin{tabular}{p{0.24\columnwidth}p{0.66\columnwidth}}
\toprule
Notation & Description \\
\midrule
$N$ & Number of questions in the dataset \\
$V$ & Number of nodes in KB \\
$P_{\mathrm{total}}$
& Number of KB paths, including 1,2,3-hops \\
$C_{\mathrm{LLM}}$ & Cost of one LLM call \\
$C_{\mathrm{enc}}$ & Cost of one text-to-embedding operation \\
$d$ & Embedding dimension \\
$N_{\mathrm{cand}}$ & Number of candidate paths for a question \\
$top_k$ & Number of candidates retained for reranking \\
\bottomrule
\end{tabular}
\end{table}

\begin{figure}[t]
        \centering
        \includegraphics[width=\linewidth]{figures/cases.pdf}
        \caption{Two cases. Highlights are detected entities for coarse-grained path retrieval for FashionEcoKG.}
        \label{fig:result_case}
\end{figure}

During preprocessing, PG-RAG generates pruned questions and extracts entities with LLM calls, incurring \(O(N \cdot C_{\mathrm{LLM}})\), and encodes questions, KB paths, and nodes with costs of \(O(N \cdot C_{\mathrm{enc}})\), \(O(P_{\mathrm{total}} \cdot C_{\mathrm{enc}})\), and \(O(V \cdot C_{\mathrm{enc}})\), respectively. Although path encoding dominates due to the large number of paths, it is performed only once during KB construction. Moreover, since \(C_{\mathrm{enc}} \approx 0.02\) seconds and encoding supports batching, this offline cost is manageable.

For online inference, each query first performs coarse retrieval through entity encoding and node similarity search, with costs \(O(e \cdot C_{\mathrm{enc}})\) and \(O(e \cdot V \cdot d)\), respectively. It then applies PSR scoring over candidate paths with cost \(O(N_{\mathrm{cand}} \cdot d)\), followed by GAR reranking of the top-\(k\) paths using LLM calls, incurring \(O(\text{top}_k \cdot C_{\mathrm{LLM}})\). Finally, answer generation requires \(O(C_{\mathrm{LLM}})\). Thus, the overall online cost is mainly governed by node similarity search and LLM reranking over a small set of top-\(k\) candidates.

In our experiments, we set $\text{top}_k=10$, which keeps the average online retrieval time of PG-RAG at approximately 1.5 seconds per query. The subsequent QA latency mainly depends on the chosen LLM, with an average inference time of approximately 5--8 seconds per query. By comparison, TOG requires about 60 seconds per query for retrieval, while KAPING takes around 1 second per query when embeddings are pre-computed. These results show that PG-RAG achieves a practical balance between retrieval effectiveness and online efficiency.

\subsection{Case Study}
\label{sec:case}
% In \autoref{fig:result_case}, we provide two illustrative examples of fashion QA, showcasing both the intermediate outputs and the final generated answers. These cases clearly demonstrate the individual and collective effectiveness of the various modules within our PG-RAG framework. 
In \autoref{fig:result_case}, we provide two illustrative examples of fashion QA, showing the original question, the pruned query, the retrieved context, and the final generated answer. These cases demonstrate how different modules in PG-RAG work together to improve retrieval reliability and reasoning accuracy.

In the first case, the original question contains multiple potentially distracting concepts, including \textit{Full Skirts}, \textit{Trench Coats}, \textit{Trench Coat Design}, and \textit{War}. Direct retrieval based on the full question may be affected by irrelevant terms such as \textit{Full Skirts}, which is not central to answering the question. Our query pruning module removes such distracting information and reformulates the question into a concise semantic skeleton: identifying the century of the war that influenced trench coat design. This pruned query better captures the core reasoning intention. As a result, the retrieval module successfully obtains \textbf{a relevant KG path} indicating that \textit{the trench coat was influenced by World War I}. The generated answer then combines the retrieved context with temporal knowledge that World War I occurred from 1914 to 1918, leading to the correct answer, i.e., the 20th century. This example shows that query pruning can reduce semantic noise and improve the alignment between the query and the relevant knowledge path.

The second case further illustrates the advantage of retrieving structured context from FashionEcoKG. The original question asks where the Hippie-influenced style of Bermuda Shorts originated, but it also contains several related fashion concepts that may confuse retrieval, such as \textit{Bermuda Shorts}, \textit{Shorts}, and \textit{Hippie style}. PG-RAG prunes the query into a more focused form that directly asks for the origin of the Hippie-influenced style. Based on this refined query, the retrieval module obtains \textbf{a compact but informative context path}: \textit{Bermuda shorts are associated with shorts inspired by Hippie style, which originated in San Francisco}. This retrieved path provides explicit evidence for the model to compare the candidate options and select the correct answer. The final response therefore grounds its reasoning in the retrieved context rather than relying only on parametric knowledge.

Overall, these examples reveal three advantages of PG-RAG. First, the pruning step improves query clarity by removing distracting or less relevant concepts while preserving the key reasoning intent. Second, the path-based retrieval over FashionEcoKG provides concise and structured evidence that connects fashion items, styles, influences, and origins. Third, the grounding and answer generation process enables the LLM to reason over the retrieved context and produce interpretable answers. Therefore, PG-RAG not only improves final QA accuracy, but also enhances transparency by exposing intermediate retrieval and reasoning results.

% \begin{figure*}[t]
%     \centering
%     % 子图 (a)
%     \begin{subfigure}{0.5\linewidth}
%         \centering
%         \includegraphics[width=\linewidth]{figures/cases.pdf}
%         \caption{Two cases. Highlights are detected entities for coarse-grained path retrieval for FashionEcoKG.}
%         \label{fig:case}
%     \end{subfigure}
%     \hskip 2mm 
%     % 子图 (b)
%     \begin{subfigure}{0.45\linewidth}
%         \centering
%         \includegraphics[width=\linewidth]{figures/qs_sample (1).pdf}
%         \caption{Fashion QA examples. Correct answer is highlighted.}
%         \label{fig:qa_example}
%     \end{subfigure}
%     % 整个大图的主标题
%     \caption{Demonstration of FashionQA cases and QA examples.}
%     \label{fig:combined_cases_and_qa}
% \end{figure*}
\section{Conclusion}
This paper addresses knowledge-intensive fashion question answering with FashionEcoKG, an expert-anchored knowledge graph built through a three-stage hierarchical pipeline that grounds academic concepts in authoritative textbooks and expands them via cross-domain KG alignment and open-world data, capturing the broader fashion ecosystem overlooked by existing resources. To leverage this structured intelligence, we propose PG-RAG, whose Dual-Granularity Path Selection module combines Pruning-based Semantic Ranking to neutralize query noise and Grounding-based Agentic Ranking to provide logical verification. PG-RAG significantly outperforms text-based and KG-RAG baselines, confirming that while LLMs hold sufficient parametric memory for basic terminology, our framework supplies the structural intelligence required for complex cross-concept synthesis in expert-intensive vertical domains such as fashion.


\end{document}